\documentclass[aps,prb,twocolumn]{revtex4-1}
\usepackage{graphicx}
\usepackage{times}
\usepackage{color}
\input pdfcolor.tex
\usepackage{graphicx}
\usepackage{epsfig,color}
\begin{document}

\title{Population imbalanced lattice fermions near the BCS-BEC crossover: \\
I.~The breached pair and metastable FFLO phases}

\author{Madhuparna Karmakar and Pinaki Majumdar}
\affiliation{Harish-Chandra Research Institute, 
Chhatnag Road, Jhunsi, Allahabad 211 019, India}

\begin{abstract}
We study s-wave superconductivity in the two dimensional attractive 
Hubbard model in an applied magnetic field, assume the extreme Pauli 
limit, and examine the role of spatial fluctuations in the coupling 
regime corresponding to BCS-BEC crossover. 
We use a decomposition of the interaction in terms of an auxiliary
pairing field, retain the static mode, and sample the pairing field
via Monte Carlo. The method requires iterative solution of the 
Bogoliubov-de-Gennes (BdG) equations for amplitude and phase
fluctuating configurations of the pairing field.
We establish the full thermal 
phase diagram of this strong coupling problem, revealing $T_c$ scales an 
order of magnitude below the mean field estimate, highlight the spontaneous 
inhomogeneity in the field induced magnetization, and discover a strong 
non monotonicity in the temperature dependence of the low energy density 
of states. We compare our results to the experimental phase diagram of 
the imbalanced Fermi gas at unitarity.  This paper focuses on the magnetized 
but homogeneous (breached pair) superconducting state, a companion paper 
deals with the Fulde-Ferrell-Larkin-Ovchinnikov (FFLO) regime.
\end{abstract}

\date{\today}
\maketitle

\section{Introduction}
 
For an electron system in a superconducting state the Meissner effect
characterizes the response to a magnetic field. In type II superconductors
 there is flux penetration beyond a threshold $h_{c1}$ 
in the form of an Abrikosov lattice
\cite{parks,bennemann}, before superconductivity (SC) is finally lost
at the `orbital critical field' $h_{c2}^{orb}$.
The magnetic field also couples to the spin of the electrons, 
and tends to break an `$\uparrow \downarrow$' pair. 
This effect is detrimental to SC, and, if
orbital effects were irrelevant, SC order would be lost at some
`Pauli limiting' field, $h_{c2}^P$, say 
\cite{clogston1962,chandrashekhar1962,shimahara2007}.
The ratio of these critical fields, $\alpha = h_{c2}^{orb}/h_{c2}^P$,
defines the Maki parameter and is roughly 
\cite{maki1964,saint-james1969} 
$\Delta_0/\epsilon_F$,
where $\Delta_0$ is the zero temperature gap in the SC state and
$\epsilon_F$ is the Fermi energy.

In most superconductors $\alpha \ll 1$, so the
Pauli suppression effects never show up. There are however three
scenarios where it becomes relevant. (a)~If $\epsilon_F$ is 
suppressed strongly by correlation effects, as in heavy fermions
where the suppression factor can be $\sim 10^3$ due to Fermi liquid
corrections \cite{stewart_rmp}, (b)~for two dimensional systems, the layered
organics, say, orbital effects are
irrelevant for an `in plane' field, and (c)~for {\it neutral} Fermi
gases, as in cold atomic systems, the magnetic effects would 
be related only to spin.
Recent discoveries on the heavy fermion 
\cite{bianchi2003,tayama2002,koutro2010,
kenzelmann2008,capan2004,martin2005,kenzelmann2014,kumagai2006,vekhter2010}
CeCoIn$_5$, the $\kappa$-BEDT based layered 
superconductors 
\cite{beyer2012,lortz2007,mayaffre2014,wright2011,coniglio2011,bergk2011,agosta2012,cho2009}, 
iron pnictides \cite{zocco2013, cho2011, khim2011, ptok2013,ptok2014} and population imbalanced cold Fermi
gases \cite{partridge2006,ketterle2008,ketterle2007,shin2006,liao2010,shin2008, capone2008,
machida2004, machida2006, sun2011}, 
make the Pauli limit relevant.

\begin{figure}[b]
\centerline{
\includegraphics[width=4.3cm,height=5.0cm,angle=0]{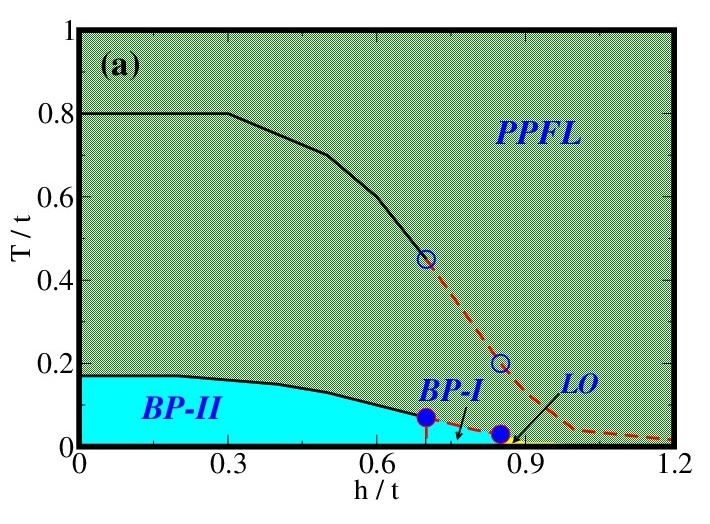}
\includegraphics[width=4.3cm,height=5.0cm,angle=0]{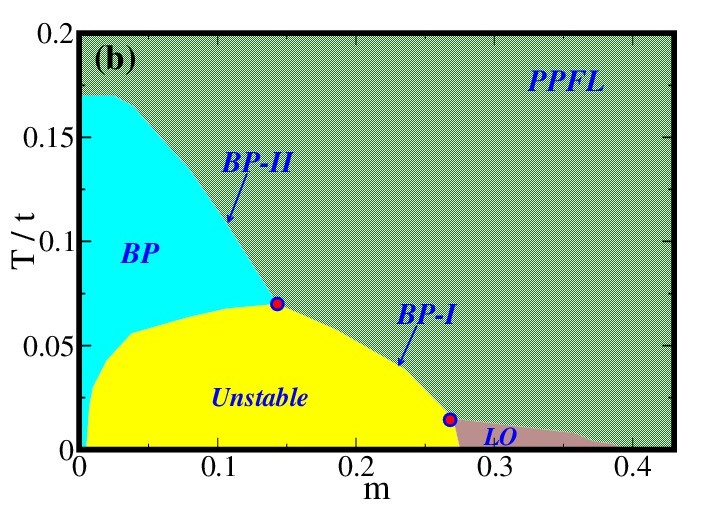}
}\label{fig1}
\caption{Color online: (a) Comparison of $T_{c}$ scales
 obtained from the mean field calculation (upper curve)
and our static auxiliary field (SAF) Monte Carlo technique
(lower curve).
In the SAF data BP-II represents a breached pair state that
undergoes a second order transition to the partially polarized
Fermi liquid (PPFL), while BP-I undergoes a first order
thermal transition to the PPFL. Beyond BP-I the system
exhibits FFLO order up to some critical field. 
(b) Polarization -vs- temperature phase diagram inferred from
the SAF calculation, plotted in the spirit of the experimentally
obtained unitary Fermi gas result \cite{ketterle2008}.
The `unstable' region is phase separated.
}
\end{figure}

Early extensions \cite{sarma1963} 
of the BCS scheme to finite Zeeman field
(neglecting orbital effects) predicted that, in the continuum,
the superconducting $T_c$ decreases with applied field up to
a critical value, $h_1$, say, and the thermal transition remains
second order. Beyond $h_1$, one would have expected a SC state
with a first order thermal transition, but the ground state
actually becomes modulated, in the spirit predicted by
Fulde and Ferrell (FF) \cite{ff} and Larkin and 
Ovchinnikov (LO) \cite{lo}, and stays so till all order
is lost at some $h_{c2}$. 
{The FFLO state is characterized by  periodic
spatial
modulation of the superconducting order parameter and 
magnetization. The modulations in these quantities are
complementary and the nodes of the 
superconducting order correspond to the maximum of the 
magnetization, and {\it vice versa}. 
}
The zero temperature and $h < h_1$ system is an `unpolarised
superfluid' (USF), its finite temperature counterpart 
is a `breached pair' (BP) state 
and the $h_1 < h < h_{c2}$ window is FFLO.
{The BP phase is
the finite temperature extension of
the USF, with 
homogeneous superconducting order spatially 
coexisting with finite uniform polarization.
}
The original scenario does not support any
first order transition between the BP phase
and the normal state.  

Experiments bear out some features of this scenario
with solid state studies focused on probing the
FFLO state while cold atom experiments probe the
general effect of imbalance on pairing.
The FFLO signatures in CeCoIn$_5$ include
specific heat \cite{bianchi2003}, magnetic torque \cite{gratens2012}, 
muon spin
relaxation \cite{spehling2009}, NMR \cite{koutro2010,kumagai2006}, 
and magnetic neutron scattering \cite{kenzelmann2014} while in
the $\kappa$-BEDT based organics there is indirect
evidence \cite{beyer2012,lortz2007,
mayaffre2014,wright2011,coniglio2011,bergk2011,agosta2012,cho2009}
for a modulated state at large in-plane fields.  
In KFe$_{2}$As$_{2}$ also thermal expansion and magnetostriction 
suggests the occurrence of Pauli limited
superconductivity \cite{zocco2013}.
For cold atoms, fermionic superfluidity
with population imbalance 
has been probed in detail with the Fermi
gas tuned to unitarity \cite{ketterle2008}
revealing an `universal' phase diagram. 

The microscopic models for superconductivity
(or superfluidity) in these systems are widely different but 
they share the features of (i)~a `homogeneous' magnetized 
superfluid state near $T_c$ at intermediate fields, (ii)~a 
possible
FFLO state at higher fields, and (iii)~being in a coupling
regime {\it well beyond the reach of mean field theory}
(at least for the atomic gases).
Taking (iii) as our point of departure we 
address these issues by studying 
the Zeeman field dependence in the 
attractive two dimensional Hubbard model
at intermediate coupling, $U/t=4$ (see later). 
This corresponds roughly to the maximum 
$T_c$ in the BCS-BEC crossover window, 
and crucially involves amplitude and phase 
fluctuations in describing the thermal physics
\cite{bloch2008, strinati2008, levin2005, tarat2014}.
Our main results, using a recently developed Monte Carlo 
(MC) approach, are the following:
\begin{enumerate}
\item
We observe that in the imbalanced problem, as in the case of balanced 
Fermi gases  
\cite{engelbrecht1993, haussmann2007, perali2004, bulgac2007, burovski2007, akkineni2007}, 
the fluctuation effects suppress $T_c$ scales 
by a factor 
of more than $4$ compared to widely used mean field theory.
\item
Intermediate fields allow for a temperature window
over which the superfluid supports significant magnetization 
which, although homogeneous on the average, shows noticeable
configurational fluctuation.
\item
At high fields the superfluid shows a first order transition
to the normal state on heating, but cooling in this field window
inevitably traps the system into a metastable FFLO state. 
\item
The spin resolved density of states shows a pseudogap (PG) 
feature that is strongly non monotonic in temperature: the pseudogap 
weakens initially with increasing temperature and then deepens again
beyond a scale $T_{max}$. The applied field dramatically suppresses
$T_{max}$.

\end{enumerate}

\setcolor\cmykBlack

We characterize the thermal state via real space maps, 
the structure
factors associated with the superfluid and magnetic 
order, the spin resolved momentum distribution
of the fermions, and the density of states. 

The rest of the paper is organized as follows. In Section-II we 
discuss 
the model and the methods used to study it.
Section-III contains the results.  Section-IV discusses possible
limitations of our numerical scheme, suggests the connection to
Ginzburg-Landau phenomenology, and relates our 
predictions to some cold atom experiments. 
Section-V concludes with our key observations.
An appendix describes the Hubbard-Stratonovich
transformation and the related approximations
in detail.

\section{Model and method}

\subsection{Model}
We study the attractive two dimensional Hubbard model 
(A2DHM) on a square lattice
in the presence of a Zeeman field:
\begin{equation}
H  = H_0  
- h \sum_{i} \sigma_{iz}  
- \vert U \vert \sum_{i}n_{i\uparrow}n_{i\downarrow} \label{eq1}
\end{equation}
with, 
$ H_0 =  \sum_{ij, \sigma}(t_{ij} - \mu \delta_{ij}) 
c_{i\sigma}^{\dagger}c_{j\sigma}$,
where 
$t_{ij} = -t$ only for nearest neighbor hopping and is zero otherwise.
$ \sigma_{iz} = (1/2)(n_{i \uparrow} - n_{i \downarrow})$.
We will set $t=1$ as the reference energy scale.
$\mu$ is the chemical potential and
$h$ is the applied magnetic field in the ${\hat z}$ direction.
$U > 0$ is the strength of on-site attraction. 
We will use $U/t=4$.

We wish to explore the physics beyond weak coupling, {\it i.e}, short 
coherence length.
This requires retaining fluctuations well beyond
mean field theory (MFT). We accomplish that as follows.
We use a `single channel' 
Hubbard-Stratonovich (HS) decomposition of the interaction 
term in terms of an auxiliary complex
scalar field $\Delta_i(\tau) = \vert \Delta_i(\tau) \vert 
e^{i \theta_i(\tau)}$.
A complete treatment of the resulting problem handles the 
full spatial and imaginary time, $(i,\tau)$, dependence of
the $\Delta$ - this is possible only within quantum Monte Carlo -
while mean field theory imposes a spatially 
periodic pattern and ignores the $\tau$ dependence.
We ignore the `time' dependence of the $\Delta$, but completely
retain the spatial dependence. This, as we shall see, makes our
method `mean field' at zero temperature, $T=0$, but retains
the crucial {\it thermal}
 fluctuations of the amplitude and phase 
of $\Delta_i$ that control $T_c$ scales, {\it etc}. We
discuss the formal structure of this approximation in
detail in the Appendix, and its 
limitations in the Discussion section.

The static $\Delta_i$ problem 
is described by the coupled effective Hamiltonian:
\begin{equation}
H_{eff}  =  H_0 - h \sum_{i} \sigma_{iz} 
+ \sum_{i}(\Delta_{i}c_{i\uparrow}^{\dagger}c_{i\downarrow}^{\dagger}
+ h.c) + H_{cl} 
\end{equation}
where $H_{cl} = 
 \sum_{i}\frac{\mid \Delta_{i} \mid^{2}}{U} $
is the stiffness cost associated with the now
`classical' auxiliary field.
The equation above indicates how the fermions see the
pairing field. The pairing field configurations 
in turn are controlled by the Boltzmann weight:
\begin{equation}
P\{{\Delta_{i}}\} \propto Tr_{c, c^{\dagger}}
e^{-\beta H_{eff}} 
\end{equation}
This is related to the free energy of the fermions  
in the configuration  
$\{\Delta_{i}\}$. For large and random $\Delta_i$ the 
trace needs to be computed
numerically. We generate the equilibrium $\{\Delta_{i}\}$ 
configurations 
by a Monte Carlo technique (see later) diagonalising the
 fermion Hamiltonian 
$H_{eff}$ for every attempted update of the auxiliary fields.

\subsection{Numerical method: Monte Carlo and variational calculation}

Mean field theory has been the standard tool for
exploring the effect of a Zeeman field on the
superconductor.
However, even though MFT may be reasonable in
capturing the ground state,
inclusion of amplitude and phase
fluctuations is essential as one moves beyond the $U/t \ll 1$ 
window. This issue has been widely discussed 
\cite{legett, nozieres, tamaki2008,  
dupuis2004, kopec2002, scaletter1989, 
trivedi1995,
allen1999, paiva2010, keller2001, capone2002,
capone2005, capone_njp2005, garg2005}
in the context of the
zero field BCS to BEC crossover.

Fluctuation effects have been found to suppress
the $T_{c}$,  compared to MFT estimates,
both at intermediate and strong coupling.
Measurements on the 3D unitary gas indicates 
\cite{zweirlein2012, jin2004} a peak
$T_{c}/E_F \sim 0.167$, 
while various theoretical estimates at unitarity 
include (a)~ a mean field based result \cite{engelbrecht1993}
suggesting $T_{c}/E_{F} \sim 0.6$ (b)~a $T$-matrix based result \cite{haussmann2007}
 suggesting
 $T_{c}/E_F \sim 0.16$,
(c)~fluctuation corrected mean field theory \cite{perali2004}, 
yielding
$T_{c}/E_F \sim 0.245$,
and (d)~Monte Carlo estimates yielding 
\cite{burovski2007,bulgac2007,akkineni2007}
$T_{c}/E_F \sim 0.15-0.25$. 
A very recent experiment on a 2D cold Fermi gas 
indicates \cite{ries2015} 
a peak $T_{c}/E_F \sim 0.16$,
while an interpolative theory estimate suggests 
$T_c/E_F \sim 0.1$.
Results on the 2D Hubbard model indicate
\cite{paiva} $T_{c}/t = 0.16$.  
Corrections beyond mean field theory, it is obvious, 
are essential for an accurate description beyond
weak coupling.

We include thermal fluctuations 
via our static auxiliary 
field (SAF) scheme, which, implemented using Monte Carlo,
can access system sizes larger than 
typical quantum Monte Carlo (QMC) calculations. 
This has several advantages: 
(i)~it provides an accurate estimate of the $T_c$,
(ii)~at high fields it helps in accessing  
spatially modulated (FFLO) paired states which may have
a large wavelength, and (iii)~it 
allows calculation of dynamical properties without the need for any
analytic continuation.

In order to make the study numerically less expensive 
the Monte Carlo is implemented using
a cluster approximation, in 
which instead of diagonalising the entire
$ L \times L$ lattice 
for each local update of the $\Delta_i$  a 
smaller cluster, of size $L_c \times L_c$,
 surrounding the update site is diagonalised. 
We mostly used $L=24$ and $L_c=6$ for the 
results in this paper.
The cluster
approximation has been extensively benchmarked, and
used successfully in the zero field case 
\cite{tarat2014}.
We will discuss the limitations of the SAF approach and
cluster based update at the end of the paper.

At zero temperature within the SAF scheme 
the energy is minimized over static
configurations of the field $\Delta_i$.
We have carried out variational calculations at
several fixed values of $\mu$, at different $h$,
exploring the following kinds of periodic configuration:
(i)~`axial stripes':  
$\Delta_{i} \sim \Delta_0 \cos(qx_i)$, and  
diagonal stripes 
$\Delta_{i} \sim \Delta_0 \cos(q(x_i + y_i))$, and  
(ii)~two dimensional 
modulations, $\Delta_{i} \sim \Delta_0 (\cos(q.x_{i})+
\cos(q.y_{i}))$, and of course (iii)~the unpolarised
superfluid (USF) state $\Delta_i = \Delta_0$.
We minimize the energy with respect to the $q$, and $\Delta_0$
(assumed real).
This paper focuses on the uniform state, the FFLO regime
is discussed in detail in a companion paper.

\subsection{Parameter regime and indicators}

Any real space numerical calculation  requires 
a system with linear dimension  
$L \gg \xi_0$, where $\xi_0$ is the $T=0$ coherence length,
to accurately capture the SC state.
Since $\xi_0$ increases with reducing $U/t$, this
puts a limit on the $U/t$ window that we can explore.
The results in this paper are at $U=4t$, both within Monte
Carlo and the variational scheme. We have also
explored $U=2t$ variationally but it requires $L \sim 48$
to access modulated phases so we have not been able to
do MC in that regime.
At $U/t=4$ we have explored the $h-T$ dependence at
multiple values of $\mu$ below half-filling (the physics
above half-filling can be inferred from this) 
but the qualitative features are independent of the choice of
$\mu$ so this paper focuses on $\mu=-0.2t$, where 
the density is $n \approx 0.94$
(independent, roughly, of $h$ or $T$). 
We have studied the
temperature dependence at a large number of fields
in the window $h/t \sim [0:1.5]$. Beyond the global
features of the $h-T$ phase diagram,
we will discuss three
field values, typical of three response regimes.

\begin{figure}[b]
\centerline{
\includegraphics[width=4.3cm,height=5.0cm,angle=0]{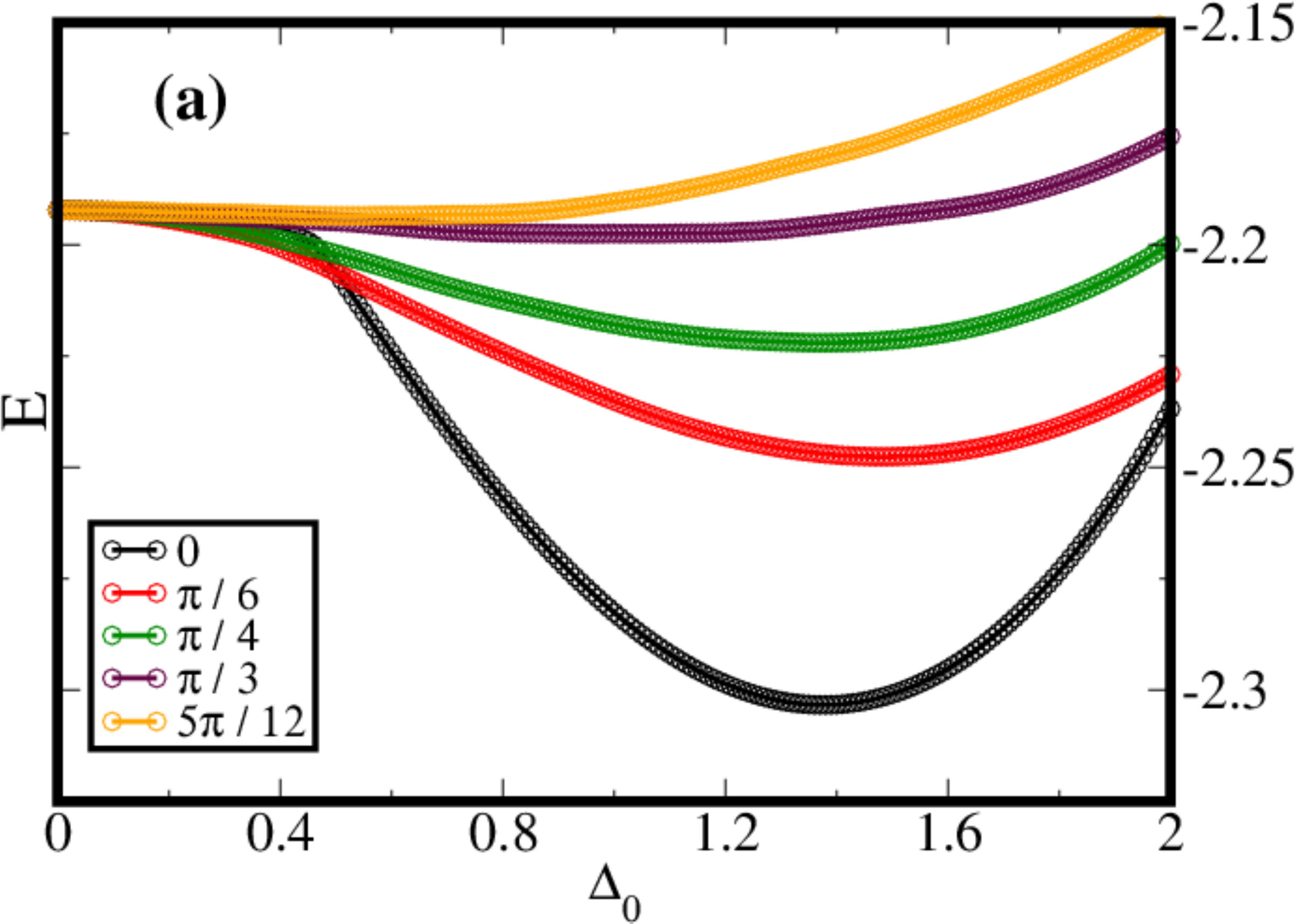}
\includegraphics[width=4.3cm,height=5.0cm,angle=0]{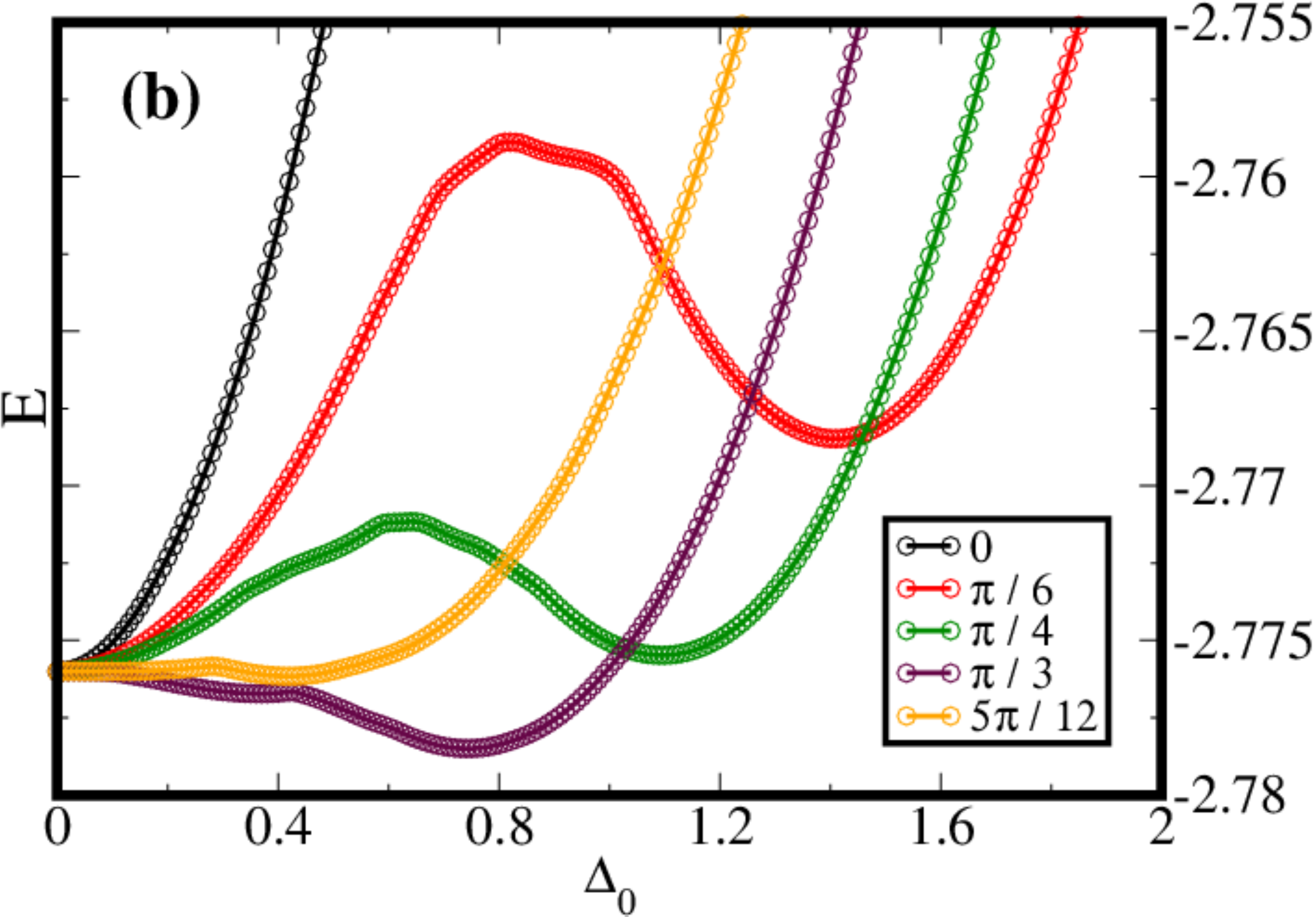}
}\label{fig2}
\caption{Color online: The variational estimate of
energy for varying amplitude $\Delta_0$ and different
modulation vectors
${\bf q} = \{2n\pi/L, 0\}$. (a) $h = 0.50t$,
where the ground state is USF, {\it i.e},
${\bf q} = (0,0)$ and (b) $h = 0.95t$ where the ground state
is amplitude modulated. We have shown data only for axial modulations,
the actual comparison is made for the full set described in the text.
}
\end{figure}
\begin{figure}[t]
\includegraphics[width=5.6cm,height=5.6cm,angle=0]{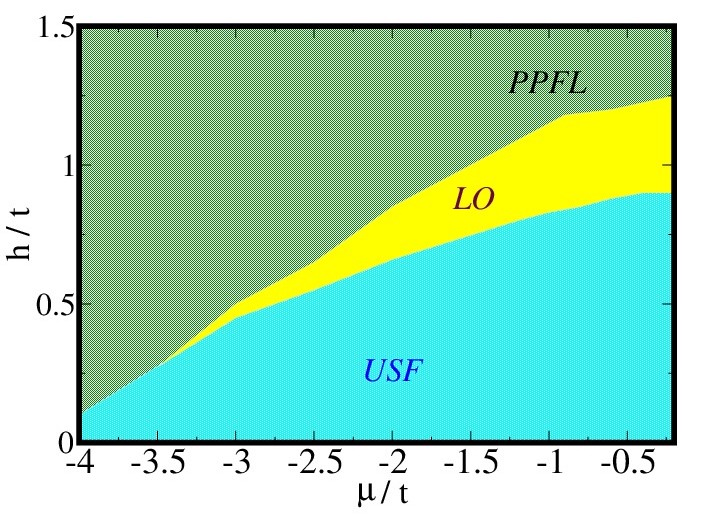}
\caption{Color online: Ground state $\mu-h$ phase diagram
obtained from the variational scheme,
showing the unpolarised superfluid (USF), modulated (LO)  and
partially polarized Fermi liquid (PPFL) regions. There is no
homogeneous superfluid state with finite magnetization,
{\it i.e}, BP, at $T=0$.}
\end{figure}

We use the following indicators to characterize the system:
(i)~Monte Carlo snapshots of $\vert \Delta_i \vert$, the
phase correlation $cos(\theta_0 - \theta_i)$ where $\theta_0$ is the
angle at a fixed reference site on the lattice, 
the magnetization variable 
$m_i = \langle n_{i\uparrow} - n_{i \downarrow} \rangle $, and
particle number 
$n_i = \langle n_{i\uparrow} + n_{i \downarrow} \rangle $.
These explicitly highlight the spatial fluctuation with
increasing temperature, and the modulated nature in the FFLO
window.
(ii)~We keep track of the structure factors, $S_{\Delta}({\bf q})$
and $S_m({\bf q})$, defined as:
\begin{eqnarray}
 S_{\Delta}({\bf q}) & = & \frac{1}{N^{2}}\sum_{i, j}
\langle \Delta_{i} \Delta_{j}^{*}\rangle
 e^{i{\bf q}.({\bf r}_{i} - {\bf r}_{j})} \cr
 S_{m}({\bf q}) & = & \frac{1}{N^{2}}\sum_{i, j}
\langle m_{i} m_{j}\rangle
 e^{i{\bf q}.({\bf r}_{i} - {\bf r}_{j})}
\nonumber
\end{eqnarray}
where, $N = L^{2}$.
(iii)~We monitor the bulk magnetization and the SC order
parameter, $S_{\Delta}({\bf q} =0; T, h)$.  
(iv)~We compute the momentum occupation
number $\langle \langle n_{{\bf k} \sigma} \rangle \rangle$
that carries the signature of imbalance and FFLO modulation.
Finally, (v)~we compute the spin resolved and total 
fermionic density of states (DOS).

\begin{figure*}[t]
\centerline{
\includegraphics[width=6.5cm,height=5.5cm,angle=0]{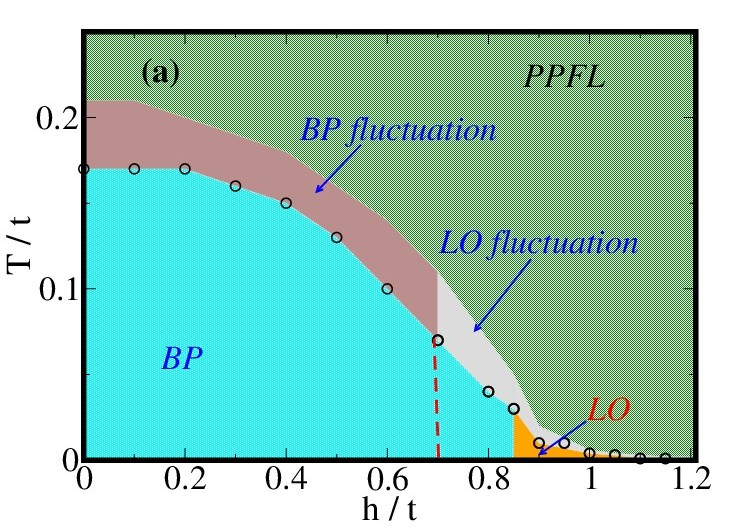}
\hspace{.3cm}
\includegraphics[width=6.5cm,height=5.5cm,angle=0]{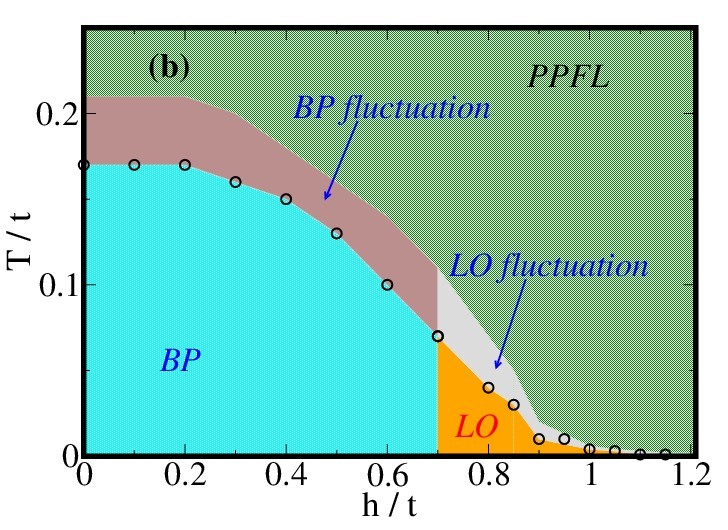}
}
\caption{Color online: The field-temperature, $h-T$,
behavior suggested by (a) heating from the variational
ground state, and (b) cooling from a random high temperature
state.  We show the phases that emerge, the $T_c$ scales, as
well as the dominant fluctuation in the disordered regime
(following a convention described in the text).
In both panels the change from a second order to first order
BP to PPFL thermal transition occurs consistently at $h \approx
0.7t$. In the heating process the first order BP-PPFL transition
encounters a region with strong LO fluctuations. This
regime shows LO fluctuations on cooling as well and the system
remains trapped in a fragmented LO state, rather than transit to
the BP phase, as $T$ is lowered. The USF to LO transition in the
ground state occurs at $h \sim 0.85t$.  The `LO' window in (a)
refers to the genuine ground state, while in (b) it also includes
the metastable LO region.  }
\end{figure*}

\section{Results}

In what follows we first highlight the 
huge difference between the mean field results 
and that of our Monte Carlo approach due to that 
of thermal fluctuations in this
coupling regime. We then take a step back to illustrate 
the working of the variational approach to the ground state
and the $\mu-h$ phase diagram that emerges. Following this
we move on to a detailed discussion of thermal properties,
in particular the difference between `cooling' and `heating'
the system, suggestive of the presence of metastable states.
We show detailed results for what we feel are three broad
field regimes: (i)~Weak field, where the $T_c$ is only
modestly modified with respect to $h=0$, the thermal transition
is second order, and there is
hardly any magnetization for $T < T_c$. (ii)~Intermediate
field, where $T_c$ is noticeably lower, the thermal transition
is still second order, but there is a window $\delta T = T_c-T >0$
where the system simultaneously shows superfluid order and
magnetization, characteristic of the `breached pair' state.
(iii)~Strong field, where the SC shows a first order
thermal transition, and there is a metastable FFLO state
over a wide temperature window. 

Fig.1 presents the primary contrast between MFT
and the MC result. Fig.1.(a) presents the $h-T$ phase diagram
indicating regions of first and second order thermal
transition and the regions of BP and
FFLO character. A much more detailed phase diagram will
be shown in Fig.4.

\begin{figure}[b]
\centerline{
\includegraphics[width=4.3cm,height=4.0cm,angle=0]{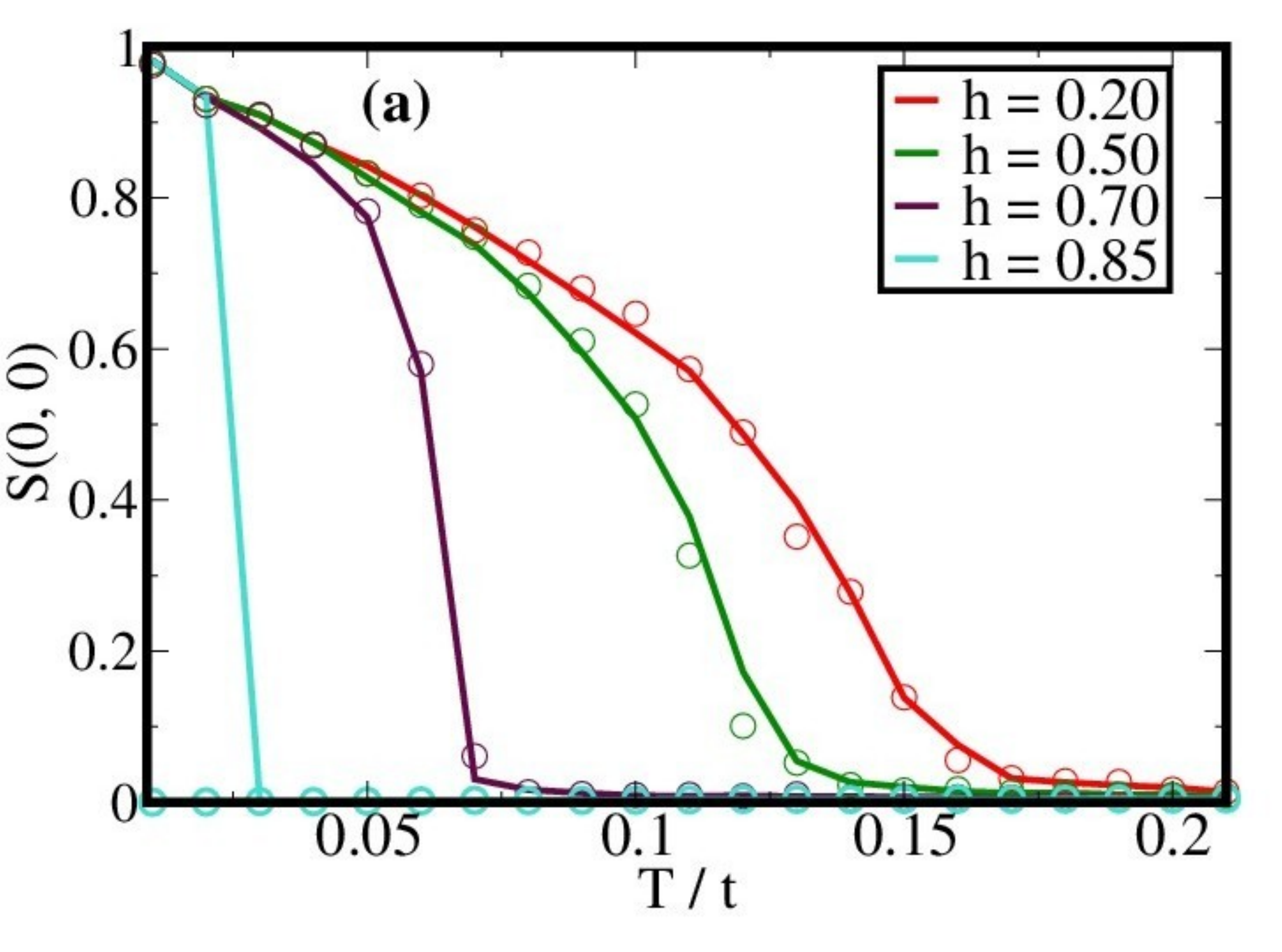}
\includegraphics[width=4.3cm,height=4.0cm,angle=0]{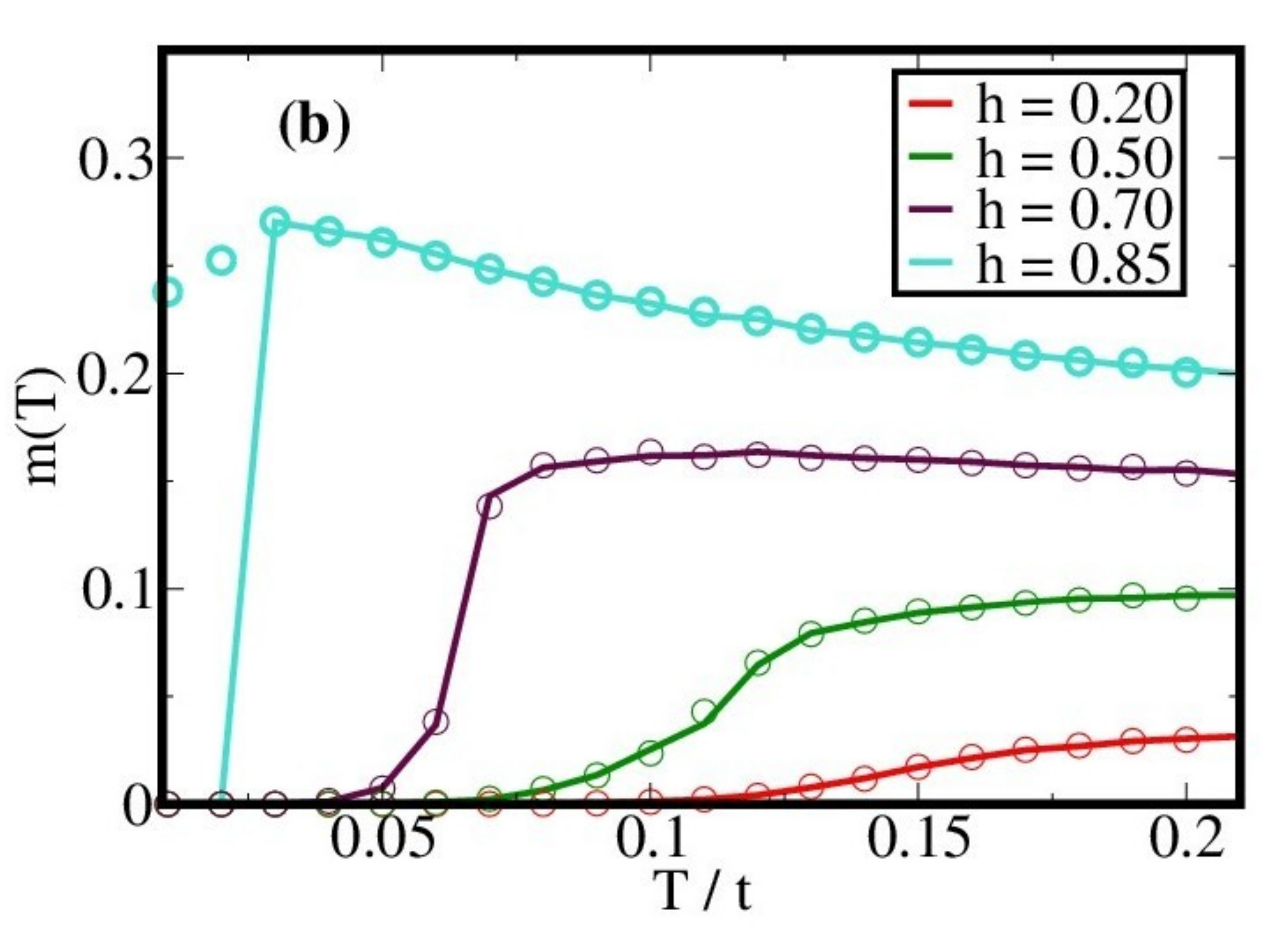}
}
\centerline{
\includegraphics[width=4.3cm,height=4.0cm,angle=0]{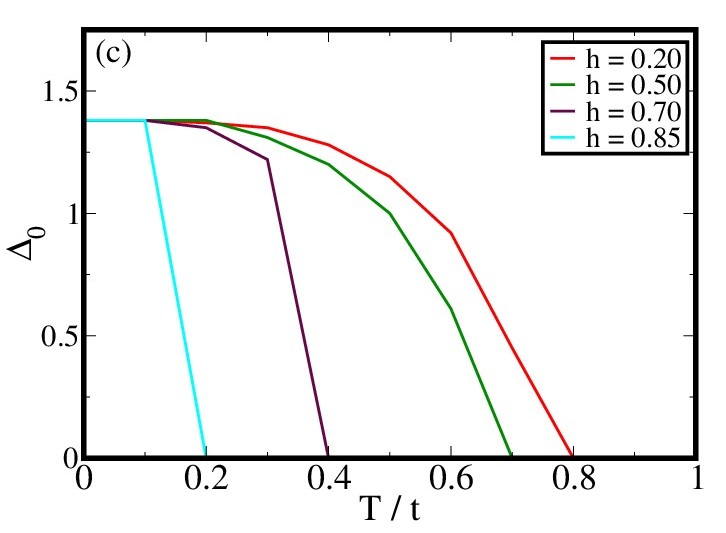}
\includegraphics[width=4.3cm,height=4.0cm,angle=0]{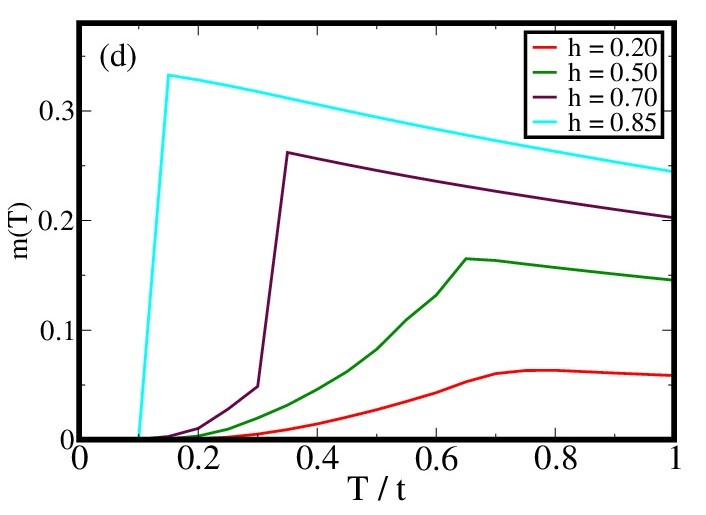}
}
\caption{Color online: (a)-(b) Monte Carlo results for the
temperature dependence of (a)~
$S(0,0)$ and (b)~the magnetization, for 
heating (solid line) and cooling (open circle).
(c)-(d).~Mean field results on the order parameter (c), 
and the magnetization (d). 
Note that the $T$ range in (c)-(d) is about 5 times larger
than in (a)-(b). Also,  the MC based order parameter shows an
almost linear drop with $T$ at low temperature while the MF 
order parameter is expectedly flat.
The magnetization results, (b)-(d), despite
their overall similarity differ in the low $h$, $T > T_c$ 
window.
}
\end{figure}

Fig.1.(b) shows the MC phase diagram in terms of the 
inferred magnetization and temperature to create a parallel
with cold fermionic systems \cite{ketterle2008}
where the physics is probed
for a fixed population imbalance (``magnetization'') 
rather than a fixed applied field.
At $T=0$ the entire USF window, $0 < h < 0.85t$, collapses
to the origin, and the first order jump to the LO state involves
a magnetization discontinuity $m \sim 0.28$. Magnetization in the LO
ground state is up to $m \sim 0.37$ beyond which we have the
`normal' partially polarized Fermi liquid. The BP state is the
finite $T$ extension of the USF and occupies a widening window
as $T$ increases and then shrinks again as $T \rightarrow T_c^0$.
The `unstable' region is the magnetization discontinuity between
the high temperature PPFL state and the low $T$ nearly 
unpolarised state
in the 1st order transition window. The LO $T_c$'s are small
and the LO phase occupies a small low temperature sliver in the
large $m$ region.
This picture helps understand the
cold atom experiments where the population imbalance, rather than
the applied field, is the primary variable. We will take up this
comparison at the end of the paper.
Fluctuations suppresses the $T_{c}$ to well below
the MF value, as observed earlier in balanced Fermi 
gases \cite{haussmann2007, perali2004, bulgac2007, burovski2007, 
akkineni2007}. The presence of imbalance (or an applied field) 
suppresses the $T_{c}$ more rapidly.

\begin{figure*}[t]
\includegraphics[width=16.0cm,height=6.4cm,angle=0]{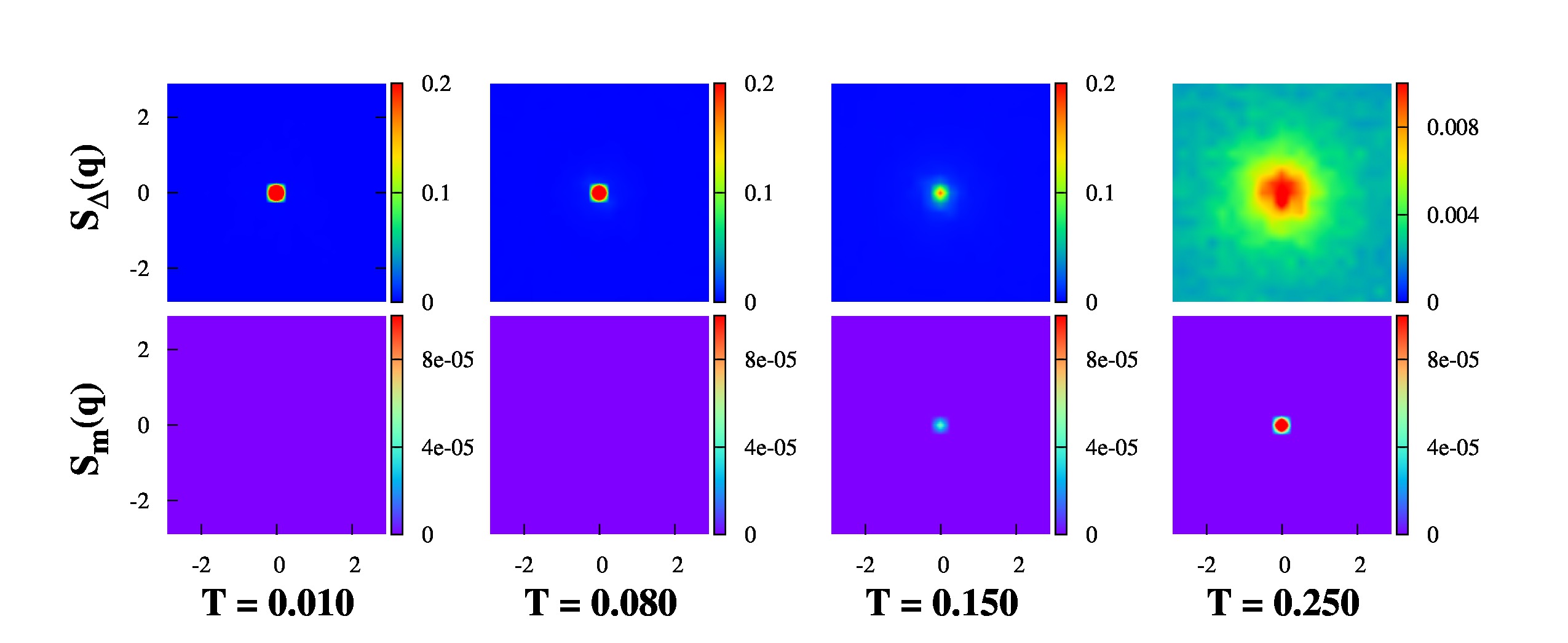}
\caption{Color online: Thermal evolution of the
superfluid ($S_{\Delta}({\bf q})$)
and magnetic ($S_{m}({\bf q})$) structure
factor at $h = 0.2t$.
By the time the magnetization picks up a reasonable value
(extreme right) the superfluid order has been lost.
}
\end{figure*}

We have used the variational approach described earlier
to determine the ground state, in the same spirit
as Chiesa {\it et al.} \cite{zhang2013,var-comment}, 
wherein diagonal, uniaxial and checkerboard
patterns of $\Delta_i$ were compared to determine 
the ground state.

Fig.2.(a) and Fig.2.(b) shows the 
dependence of the energy on the `magnitude' $\Delta_0$, 
of the pairing field,
for several values of ${\bf q}$. 
Panel (a) is for intermediate
field, $h = 0.5t$, where the ground state is still
homogeneous, {\it i.e}, at ${\bf q} =(0,0)$. 
Panel (b), at $h=0.95t$ shows
an absolute minimum at ${\bf q}
= (\pi/3, 0)$, an axial Larkin-Ovchinnikov state.

The variationally determined  $\mu-h$ 
phase diagram is shown in Fig.3. At low $h$
the system is a
homogeneous unmagnetised superfluid (USF).
One may have expected \cite{clogston1962}
this to undergo a transition 
to a partially polarized Fermi liquid (PPFL)
at a field 
$h_c = \Delta_{0}/\sqrt(2)$, the naive Pauli limit.
However, as predicted by 
Fulde and Ferrell \cite{ff} and Larkin and Ovchinnikov \cite{lo},
and confirmed by several later studies,
we find that a $\Delta_i$ modulated state 
with finite magnetization intervenes between the
USF and the PPFL.
We designate the USF to LO transition as $h_{c1}$ and
the LO to PPFL transition as $h_{c2}$. Both these
fields increase with $\mu$.
We will discuss the detailed behavior within the LO
window elsewhere.

\begin{figure*}[t]
\includegraphics[width=13.5cm,height=13.6cm,angle=0]{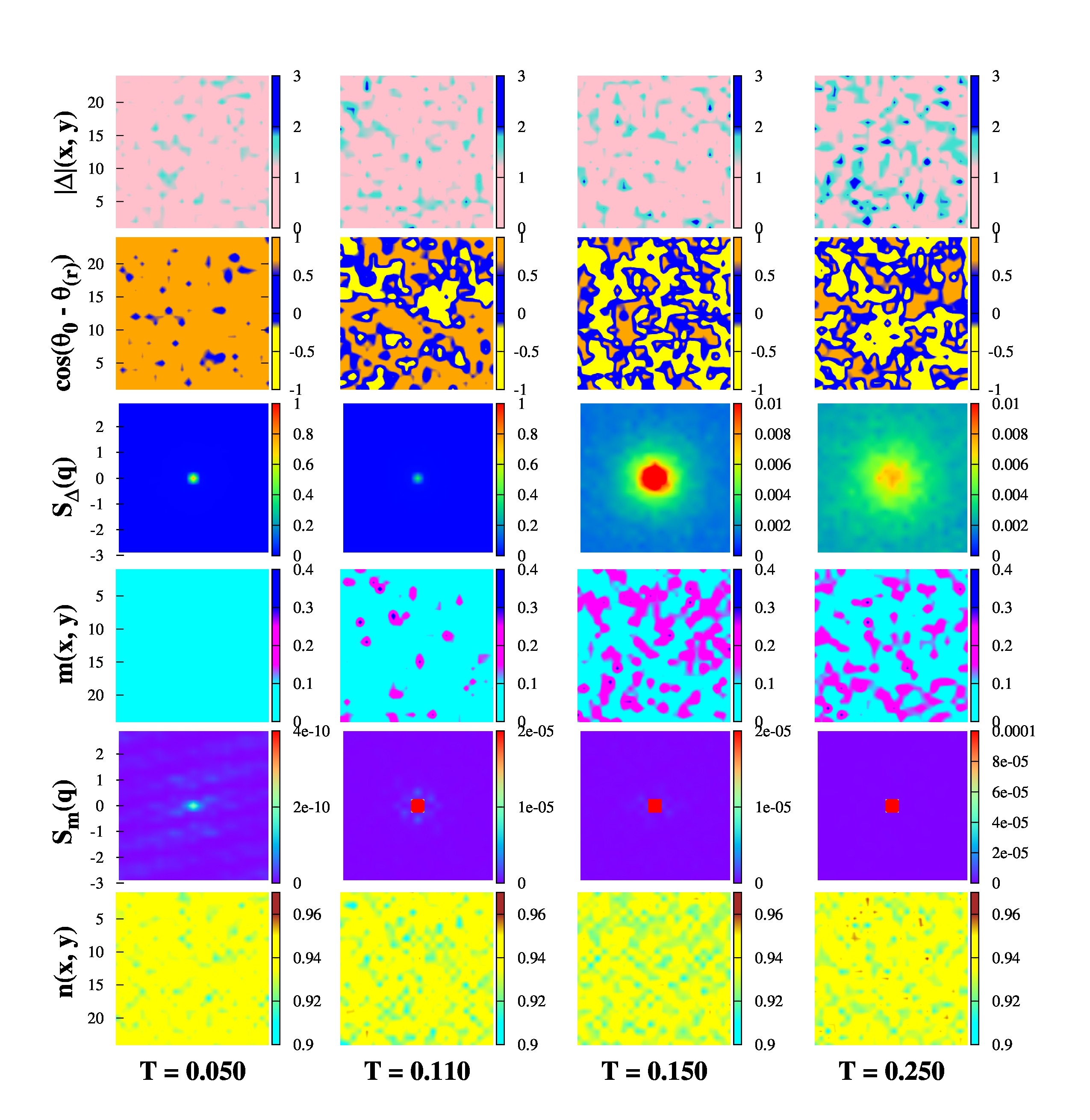}
\caption{Color online: Thermal evolution of the various
indicators at $h = 0.5t$.
Starting from the top we show maps of
$\vert \Delta \vert$, phase correlation, pairing structure factor,
magnetization $m_i$, magnetic structure factor, and number density.
The temperature, along the row, is marked at the bottom of the figure.
Between $T = 0.10t$ and $0.13t$, see Fig.5, there is both significant
superfluid order as well as magnetization.
}
\end{figure*}
\begin{figure}[b]
\includegraphics[width=8.0cm,height=4.2cm,angle=0]{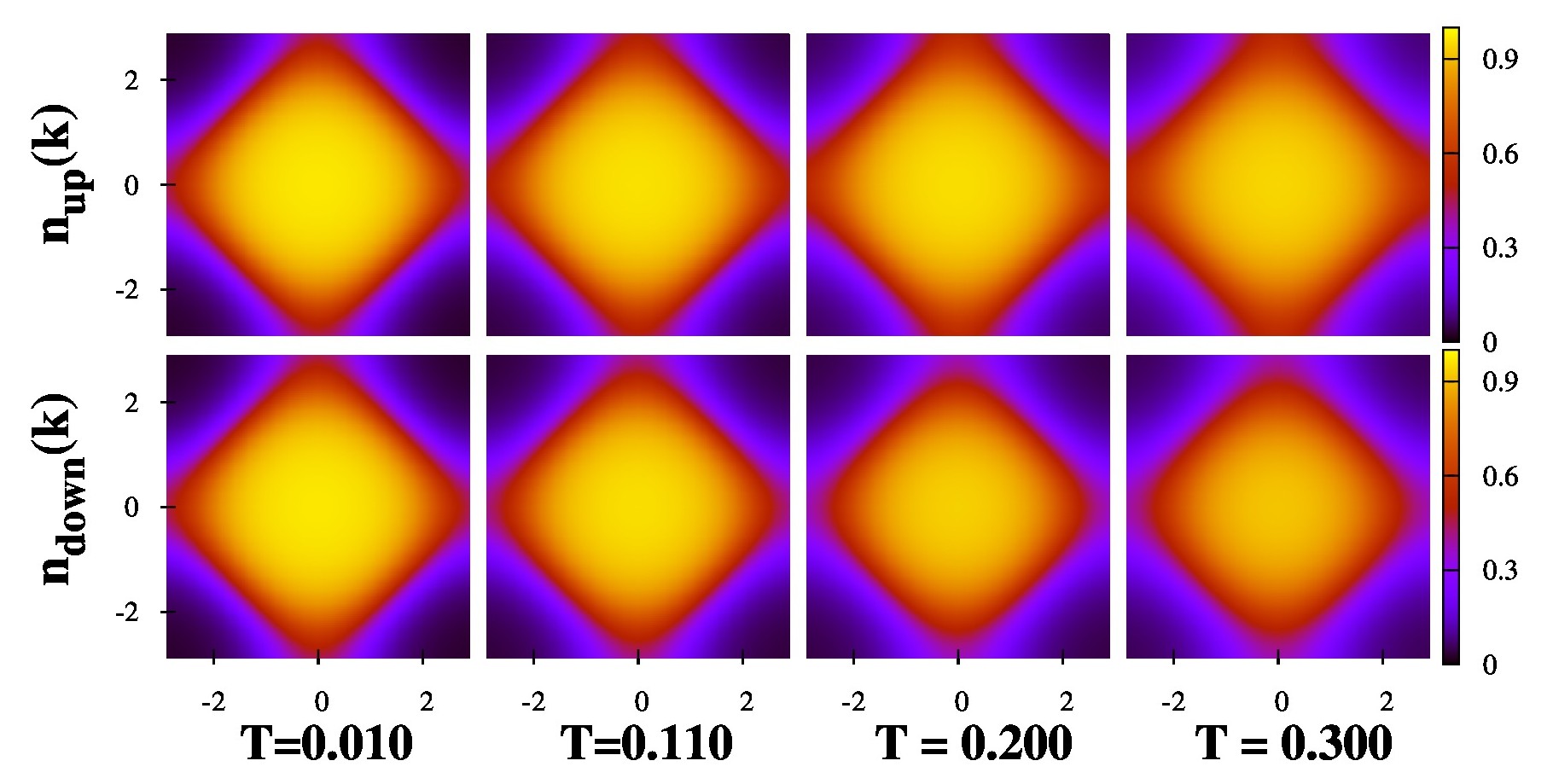}
\caption{Color online: Thermal evolution of the momentum
occupation number $n_{\sigma}({\bf k})$ at $h = 0.50$.
The up and down spin distributions are
same at $T=0$, slightly different for
$T \lesssim T_c$ (2nd row), and noticeably different
at high temperature.
} \end{figure}

\subsection{Overview of thermal phase diagram}

Mean field theory for $s$-wave superconductors in a magnetic
field indicate that (in the continuum case) the normal
to SC thermal 
transition continues to be second order from $h=0$ 
to a finite field, beyond which the system shows a first 
order transition, but now to a modulated superfluid phase
\cite{ff,lo}.

The simultaneity of the second to first order change 
and transition from the ${\bf q} =(0,0)$ to a finite
${\bf q}$ state is probably specific to continuum mean field
theory.
Additionally, the MF prediction of $T_c$ scales, {\it etc},
is valid only in the weak coupling limit.

In the presence of an underlying lattice, even MFT
suggests a field window over which one can have a 
first order SC to normal transition, see Fig.1, although the
transition temperature is badly overestimated. Beyond another
higher field the lattice based MFT predicts a modulated state. 

Fig.4 shows the phases revealed by heating from the mean field
ground state (left) and cooling (right) from a disordered
high temperature state.
The thermal transition from the SC to normal state
is second order up to a field $h_1 \sim 0.7t$ beyond
which it becomes first order (with the ordered state
still being at ${\bf q}=(0,0)$). 
For $h < 0.7t$ the results are path independent but for
$0.7t < h < 0.85t$ the system gets trapped in a LO state on
cooling although the ground state is still USF.
Beyond $h \sim 0.85t$, where the ground state is
LO the results are again path independent.

In the first order transition window,  $h_1 < h < h_{c1}$, 
the USF ground state thermally evolves into BP
at finite $T$ and then shows a transition to a PPFL
state where the fluctuations, surprisingly, have LO
character. 
On cooling down 
from a disordered state the system fails to attain 
a ${\bf q} = (0,0)$ state and instead shows strong
LO signatures. This MC inferred LO state is
energetically higher than the variational USF state
so this is a sign of metastability.
We would characterize this state in terms of the 
various indicators in a later section.

\subsubsection{Fluctuation regime}

While long range order is only
observed for $T \lesssim 0.2t$, we wanted to probe if
there is a significant window above $T_c$ where
fluctuation effects of ${\bf q} = (0,0)$ or finite
${\bf q}$ pairing can be seen.
We define the cut off to the fluctuation regime as the
temperature at which the ratio between the highest
magnitude of the structure factor peak to that
at the neighboring k-point is $\approx 1.5$.
The regimes of strong BP fluctuation and strong
LO fluctuation are marked in Fig.4 in this spirit.

\subsubsection{Thermodynamic properties}

Fig.5 shows the thermal evolution of 
${\bf q} = (0,0)$ structure factor peak, $S(0, 0)$,
 and magnetization
$m(T)$ for the magnetic fields characteristic of the low,
 intermediate and high field regimes. 
The two panels on top show the MC based results while
the lower panels are based on MFT. The MC and
MFT results have a gross similarity but
(i)~the MF $T_c$ scales are four times larger, 
(ii)~even on a normalized, $T/T_c$ scale,
the MC order parameter shows a quicker drop with temperature 
associated with the $O(2)$ nature of the superconducting
problem, while the MF plot is much flatter due to absence
of phase fluctuations, (iii)~the MF magnetization has $dm/dT <0$
above $T_c$, while the MC result, up to intermediate field,
clearly shows $dm/dT >0$.
 
The MC data also reveal that 
in the high field region, corresponding
to a first order transition,
the USF ground state is not recovered 
on cooling and the response becomes history dependent.
If one heats from the USF ground state to the $T > T_c$ and
cools again to $T=0$ a  large magnetization  state is obtained!
This state, as we will see, has prominent
LO correlations.

In what follows we provide a detailed description of the
thermal response of the imbalanced 
superconductor for three
typical field regimes.

\subsection{Low field response: the unpolarised superfluid}

We characterize $0 < h < 0.3t$ roughly as the `low field'
regime. The $T_c$ is still within $10\%$ of the $h=0$ 
value and the magnetization even
near $T_c$ is quite small ($m(T_c) \sim 0.02$ at $h=0.3t$).
As representative of this regime we
show data for  $h = 0.2t$ in  Fig.6, where 
the upper panel 
is the  pairing structure factor $S_{\Delta}({\bf q})$
and the lower panel is the 
magnetic structure factor $S_{m}({\bf q})$.
$S_{\Delta}$ loses its ordering feature 
at $T \sim 0.16t$ where the $S_m(0,0)$ 
still remains $\lesssim 10^{-5}$.

There are no finite ${\bf q}$ features in the magnetic
structure factor.
The thermal transition is reversible and no thermal history
effects show up.

\subsection{Intermediate field: breached pair state}

Next we consider the intermediate field regime
of $0.3t < h < 0.7t$. Over this window $T_c(h)$ 
falls significantly 
and the magnetization below $T_c$ reaches $\sim 0.15$ (at
$h=0.7t$). Fig.7 shows spatial features at 
$h = 0.5t$. While the ground state is still a homogeneous USF
the increase in $T$ leads quickly to development of finite
magnetization.  `Unpaired' fermions coexist with a 
${\bf q} = (0,0)$ condensate. This is a breached pair state. 

\begin{figure*}[t]
\includegraphics[width=16cm,height=18cm,angle=0]{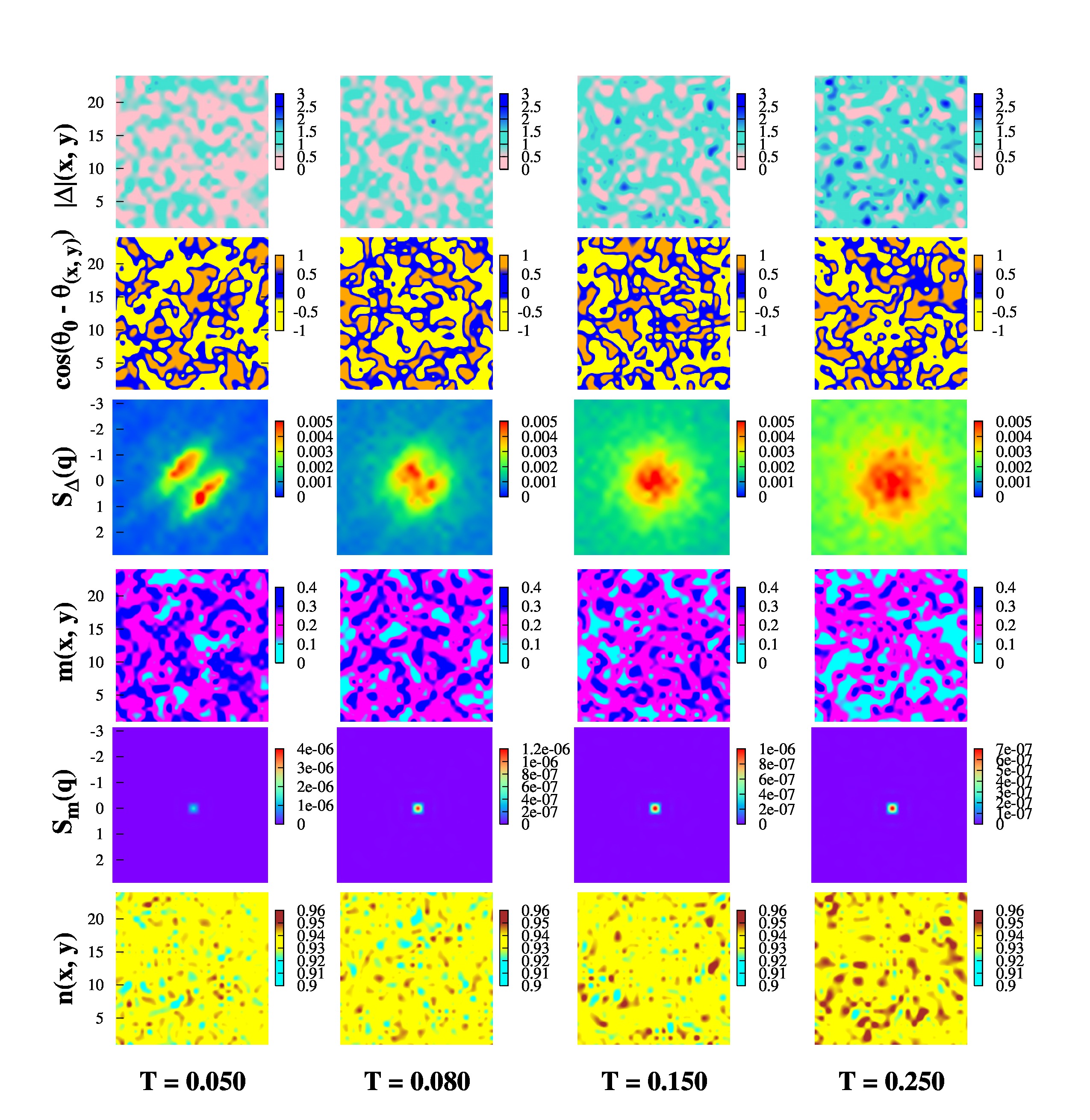}
\caption{Color online: Thermal evolution of the various
 indicators on cooling the system at $h = 0.8t$.
From top to bottom we have plotted the spatial map
of $\vert \Delta \vert$, phase correlation, the 
pairing structure factor,
magnetization, magnetic structure factor and
number density.
This illustrates the emergence of a strong LO signature
in the pairing structure factor, when the ground state is
actually USF. The pattern at the lowest temperature: $T =0.001t$
is shown later.
}
\end{figure*}

We characterize this phase in Fig.7 
through the $T$ dependence of the
following 
indicators:
(a)~the  pairing amplitude $\vert \Delta(x,y) \vert $,
(b)~phase correlation $\cos(\theta_{0}-\theta_{x,y})$ where $\theta_0$ is
the phase at a reference site,
(c)~pairing structure factor $S_{\Delta}({\bf q})$, 
(d)~magnetization $m(x, y)$, 
(e)~magnetic structure factor $S_{m}({\bf q})$,
and (f)~number density $n(x, y)$.
(a), (b), (d) and (f) are for a single MC snapshot, while (c) and (e)
are thermally averaged.
The $T_c$ in this case is $\sim 0.13t$.

With increase in temperature MC snapshots
 indicate that the $\vert 
\Delta(x,y) \vert$ becomes inhomogeneous (although a thermal
average would be homogeneous again), and the phases begin to
decohere. The $T$ dependence of  $\vert \Delta \vert$,
phase correlations, and the SC structure factor are
not qualitatively different from what we see at weak field
but $m(x,y)$ shows a departure. 
Between $T=0.11t$ and $0.15t$, {\it i.e}, across $T_c$,
we observe the emergence of 
significant magnetization in `clumps'. The magnetization, crudely,
 follows a pattern that is spatially complementary to the SC order.
The local magnetization can reach a value $\sim 0.4$
 even for $T < T_c$
(the system average however is much smaller). 
The 5th row shows the
magnetic structure factor, essentially a diffuse peak around
${\bf q} = (0,0)$, while the last row shows the density profile
(almost homogeneous).

We have calculated the momentum occupation number  
$n_{\sigma}({\bf k}) = \langle \langle c_{{\bf k}\sigma}^{\dagger}
c_{{\bf k}\sigma}\rangle \rangle$. 
In Fig.8 
we  show $n_{\uparrow}({\bf k})$ and 
$n_{\downarrow}({\bf k})$ at $h = 0.5t$ for different
temperatures. At  low temperature where the system 
is unpolarised the Fermi surfaces are of equal sizes.
As one increase the temperature the system develops 
an imbalance in the population of the up and down
fermionic species, the signature of which is observed 
in the increasing size mismatch between the two
Fermi surfaces. There is already a weak signature at
$T=0.11t$, a clear signature at  $T \sim 0.13t \sim T_c$ 
(not shown here), and a prominent difference at $T=0.2t$
and $T=0.3t$.

\subsection{High field: appearance of metastable FFLO states}

In the high field regime, $0.7t < h < 0.85t$, 
the region of first order USF to normal transition, 
the system 
seems to encounter competing minima in the energy landscape.
The state we obtain depends on the thermal history of the 
system.
We highlight the effects at a typical field $h=0.8t$.

Fig.9  shows the standard spatial and Fourier space 
indicators on a {\it cooling run}.  The results on heating from the
USF are qualitatively similar to what we have seen at $h=0.5t$.
On cooling from high $T$ the system encounters
${\bf q} \neq 0$ fluctuations and instead of transiting
to a ${\bf q} = (0,0)$ low $T$ state it actually enters a 
modulated state!
This state has higher energy than the variational USF state
which suggests its metastable character.
We show the 
real space patterns at the lowest $T$ further on.
While real space features are not very illuminating down to
$T=0.05t$, the pairing structure factor shows a clear finite
${\bf q}$ feature. 
The spatial character becomes clearer at even lower $T$ as we
show below.

\begin{figure*}[t]
\includegraphics[width=12.0cm,height=8.0cm,angle=0]{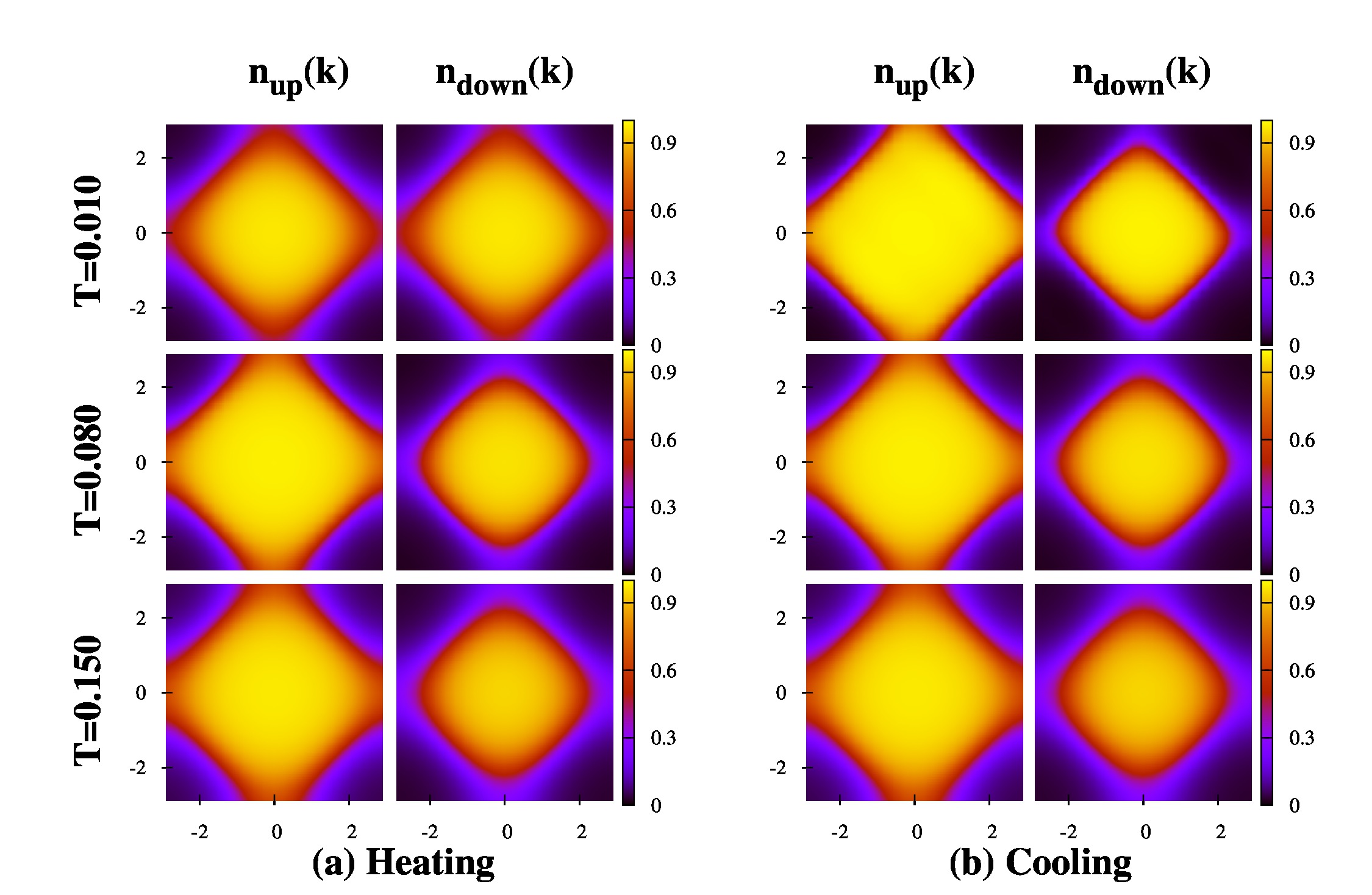}
\caption{Color online: Momentum occupation 
numbers $n_{\sigma}({\bf k})$
at different temperatures through the heating 
and the cooling cycle computed
at $h = 0.8t$. Notice the `size difference' persisting to low
temperature in the cooling run - suggesting a finite $m$ `ground
state' (actually metastable in this case). For $T > T_c$ the heating
and cooling results are essentially similar.
}
\end{figure*}
\begin{figure*}[t]
\includegraphics[width=14.0cm,height=4.0cm]{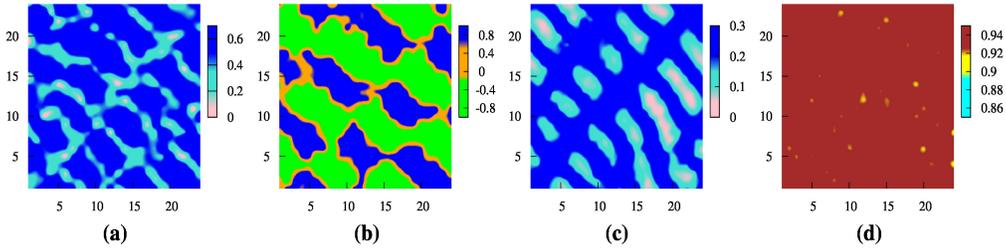}
\caption{Color online: Spatial maps characterizing the
metastable LO state through (a)~$\vert \Delta_i \vert$,
(b)~phase correlation, (c)~magnetization and
(d)~number density distribution at $T=0.001t$.}
\end{figure*}

We compute the momentum occupation numbers 
for the up and down fermionic species through the heating and
cooling cycles. Apart from the evolution of the 
mismatch between the up 
and down distributions with temperature one 
can also see the modification in 
the Fermi surface shape with respect to what one would expect in
the simple tight binding case. The straight Fermi 
surface segments  at low $T$ in Fig.10 
result from line-like LO correlations as
we show in the Fig.11.
The rise in temperature 
wipes out this feature.

What does this metastable LO state look like in real space?
We computed  the amplitude, phase, magnetization
and number density maps for MC snapshots and show a typical set
at low temperature in Fig.11.
As can be seen, real space periodic modulations are 
observed in both the superfluid order parameter 
and local magnetization.
The order parameter exhibits a nodal, domain 
wall like structure, in the 
nodes of which reside the unpaired fermions 
giving rise to a finite 
magnetization. A `node'  in the $\vert \Delta_i \vert$
corresponds roughly to a peak in the local magnetization.

\begin{figure}[b]
\includegraphics[width=7.0cm,height=3.9cm]{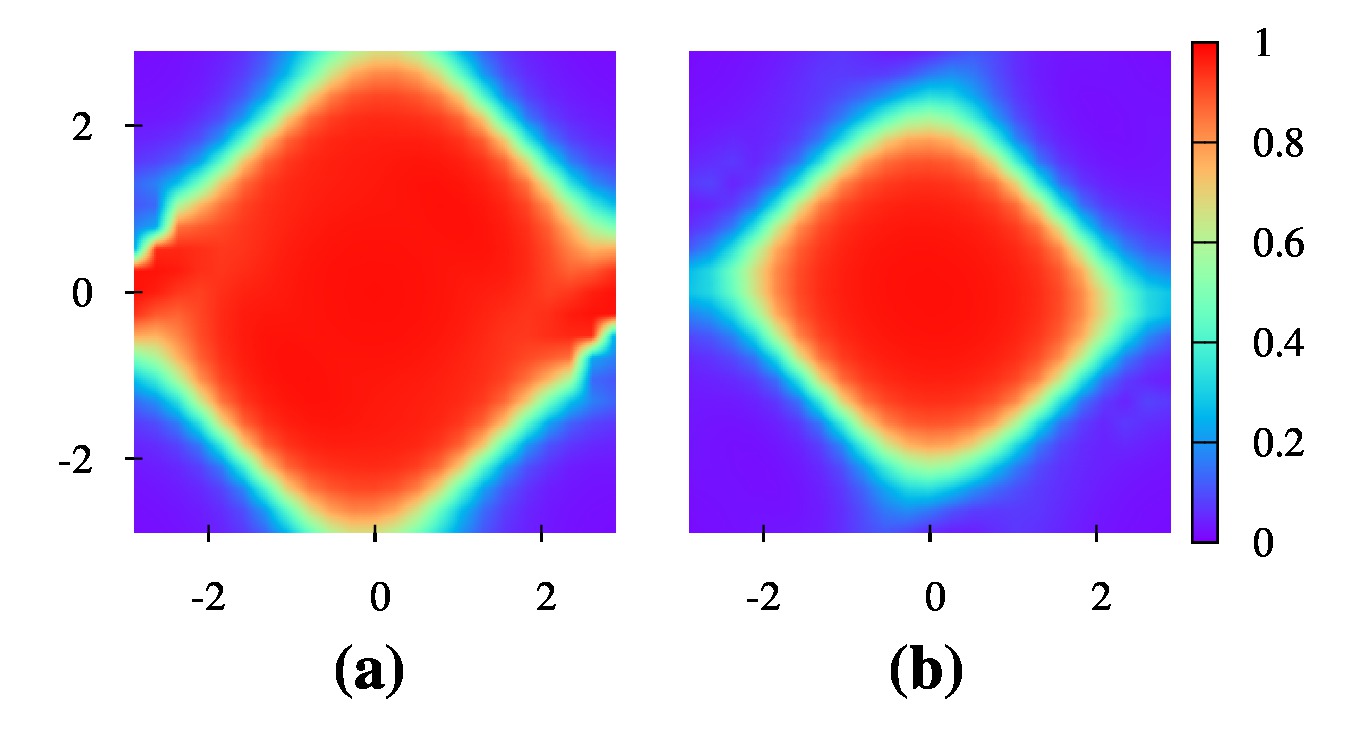}
\caption{Color online: Momentum occupation
 function for an ideal diagonal stripe phase to mimic the
pattern that we observe in Fig.11.  
Fig.11 modulations have a 2D character
(rather than simple diagonal stripe) so the actual $n({\bf k})$
in Fig.10 has an approximate fourfold look.}
\end{figure}

Before we end this section we show the ideal momentum occupation number
$n_{\sigma}({\bf k})$ corresponding to the metastable state at $h = 0.80t$
in Fig.12.  
A weaker variant of the same has been observed and presented in Fig.10.
Fig.12 prominently 
shows the anisotropic deformation of the Fermi surfaces 
in presence of an underlying modulated pairing order.

\begin{figure*}[t]
\centerline{
\includegraphics[width=5.6cm,height=5.0cm,angle=0]{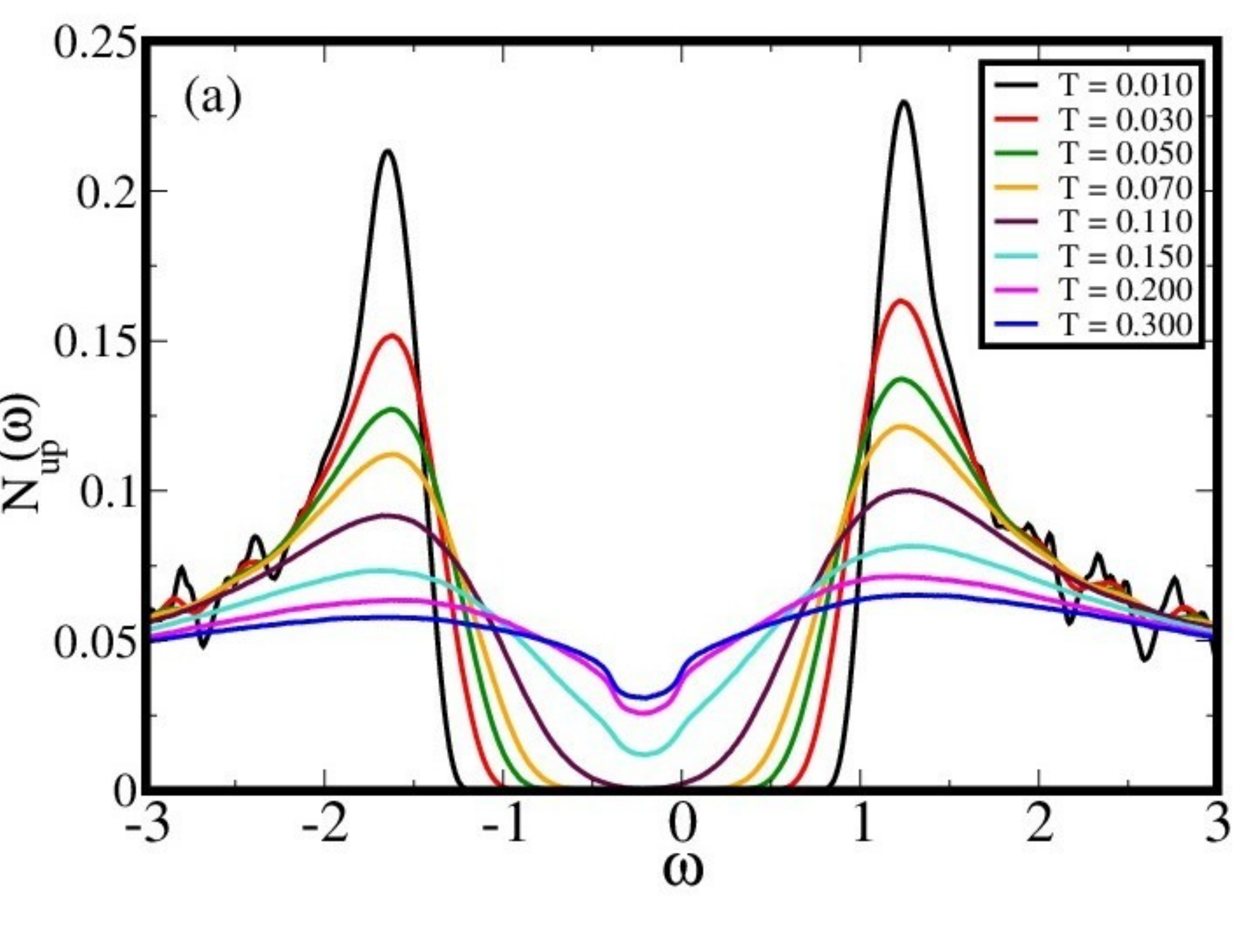}
\includegraphics[width=5.6cm,height=5.0cm,angle=0]{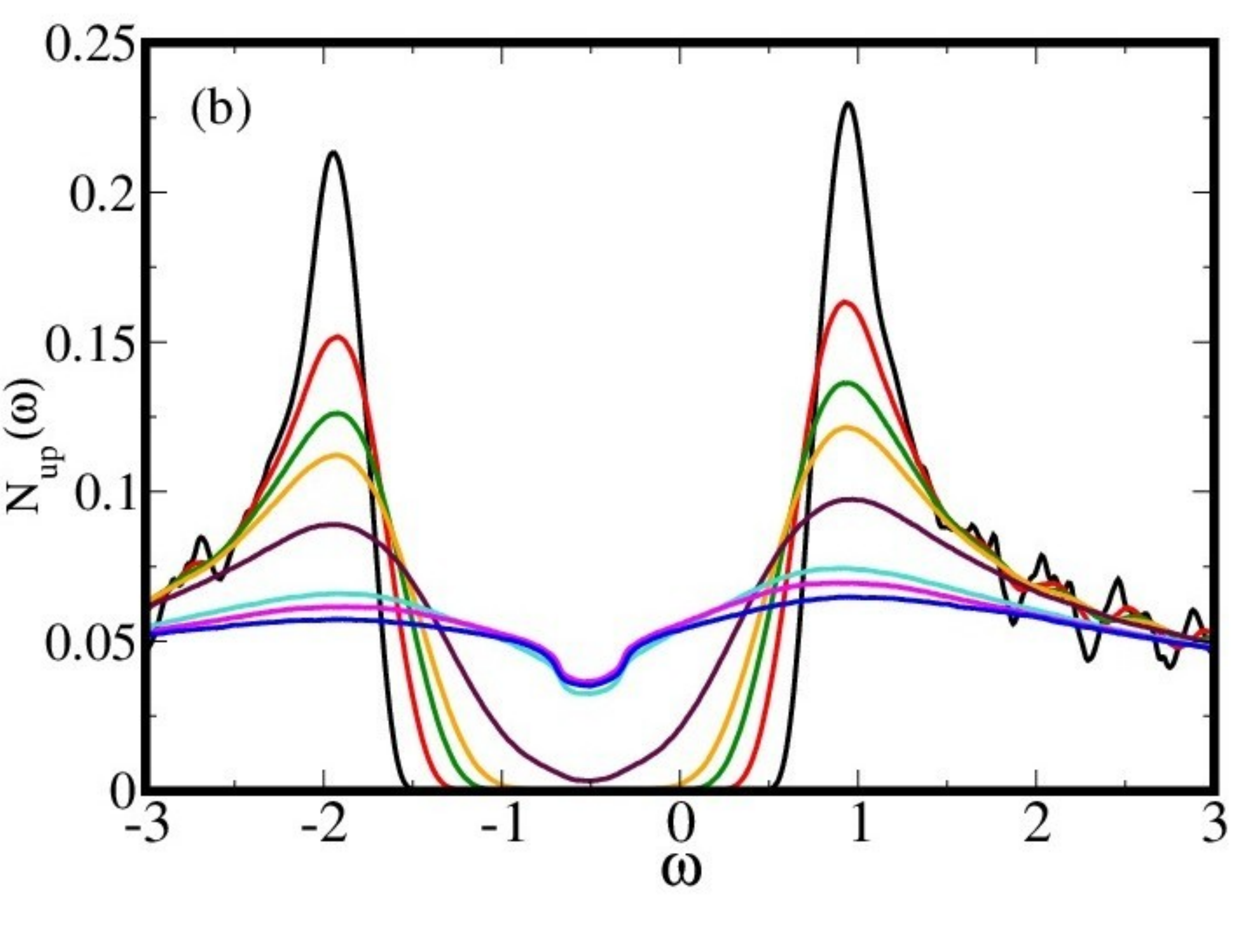}
\includegraphics[width=5.6cm,height=5.0cm,angle=0]{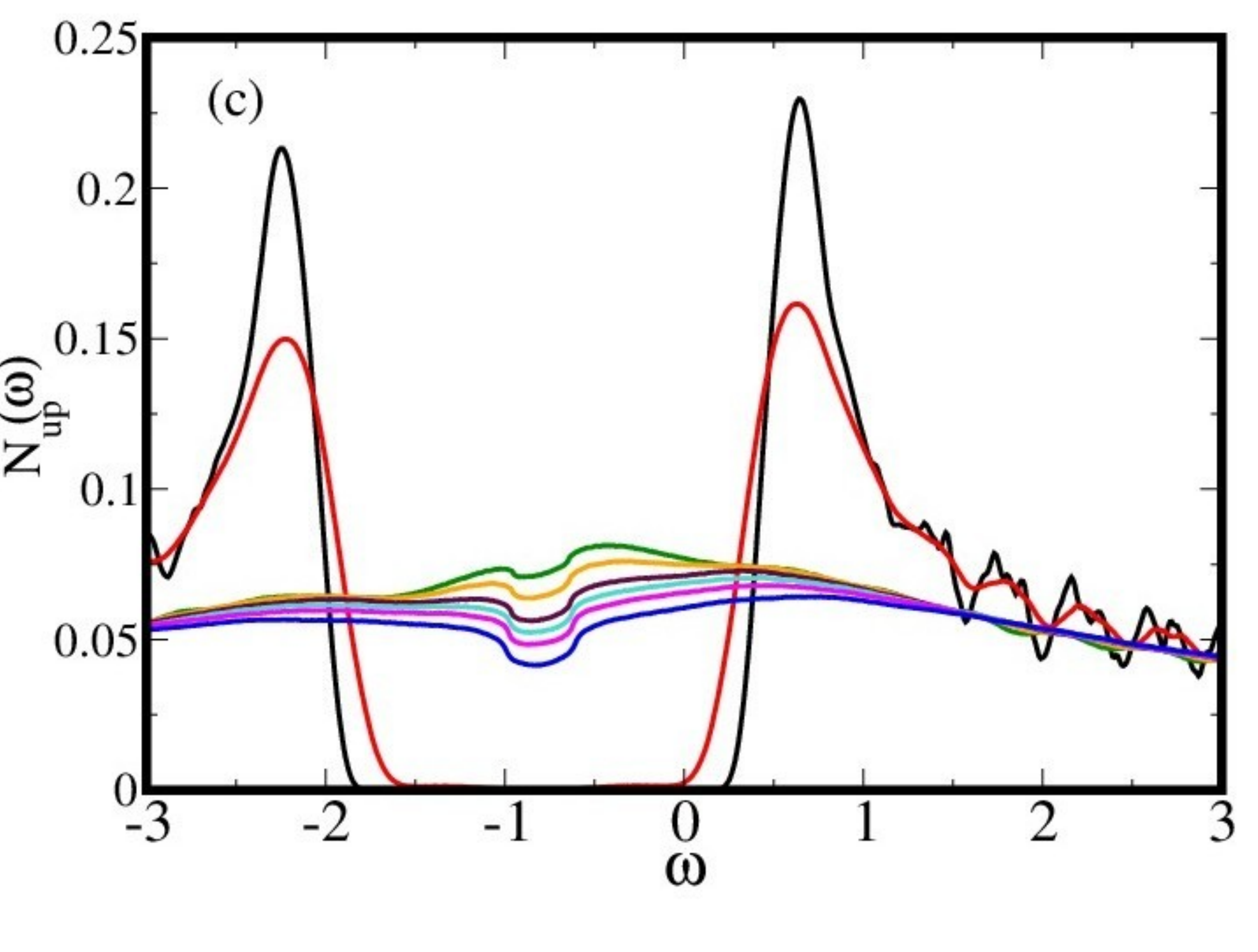}
}
\vspace{.3cm}
\centerline{
\includegraphics[width=5.6cm,height=5.0cm,angle=0]{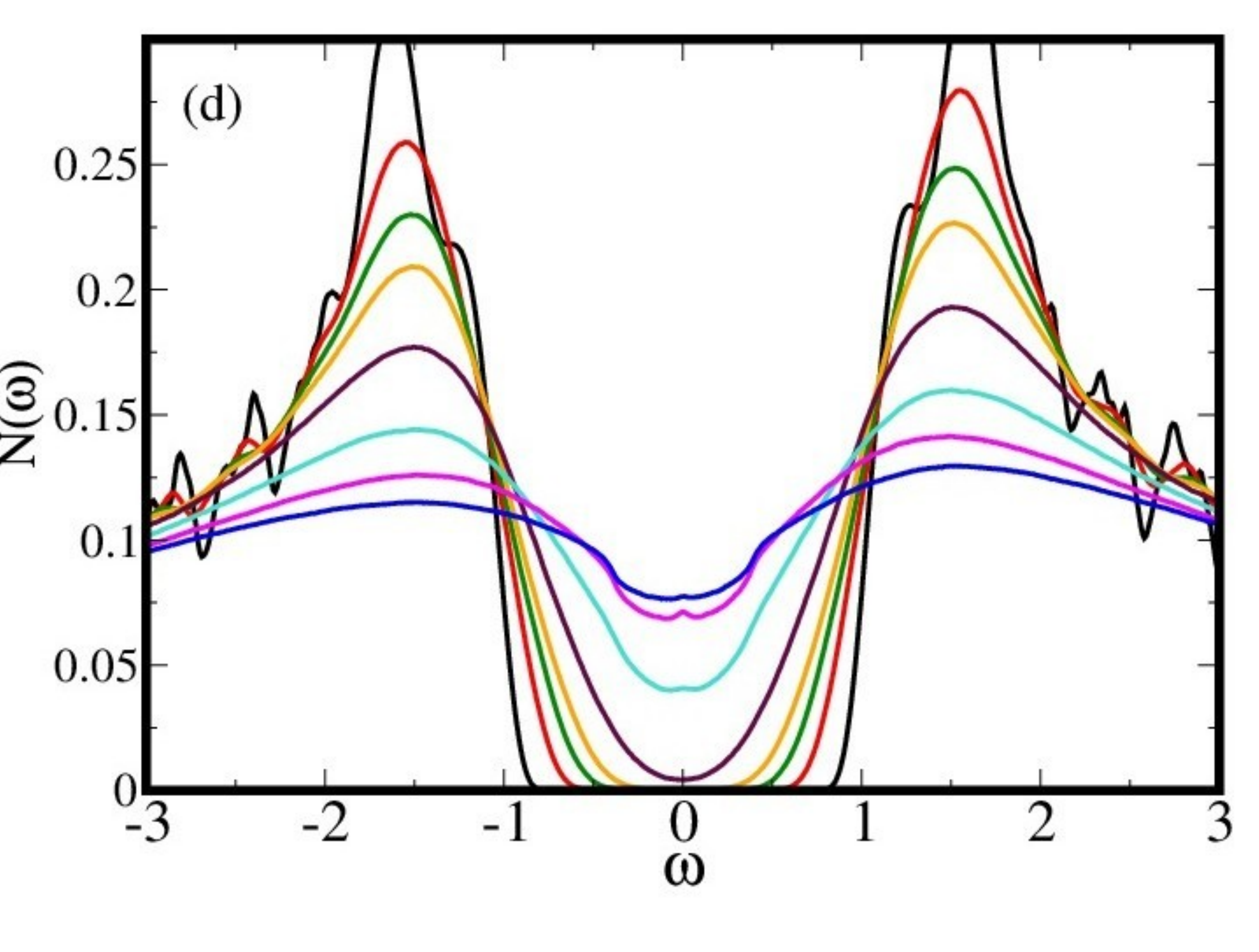}
\includegraphics[width=5.6cm,height=5.0cm,angle=0]{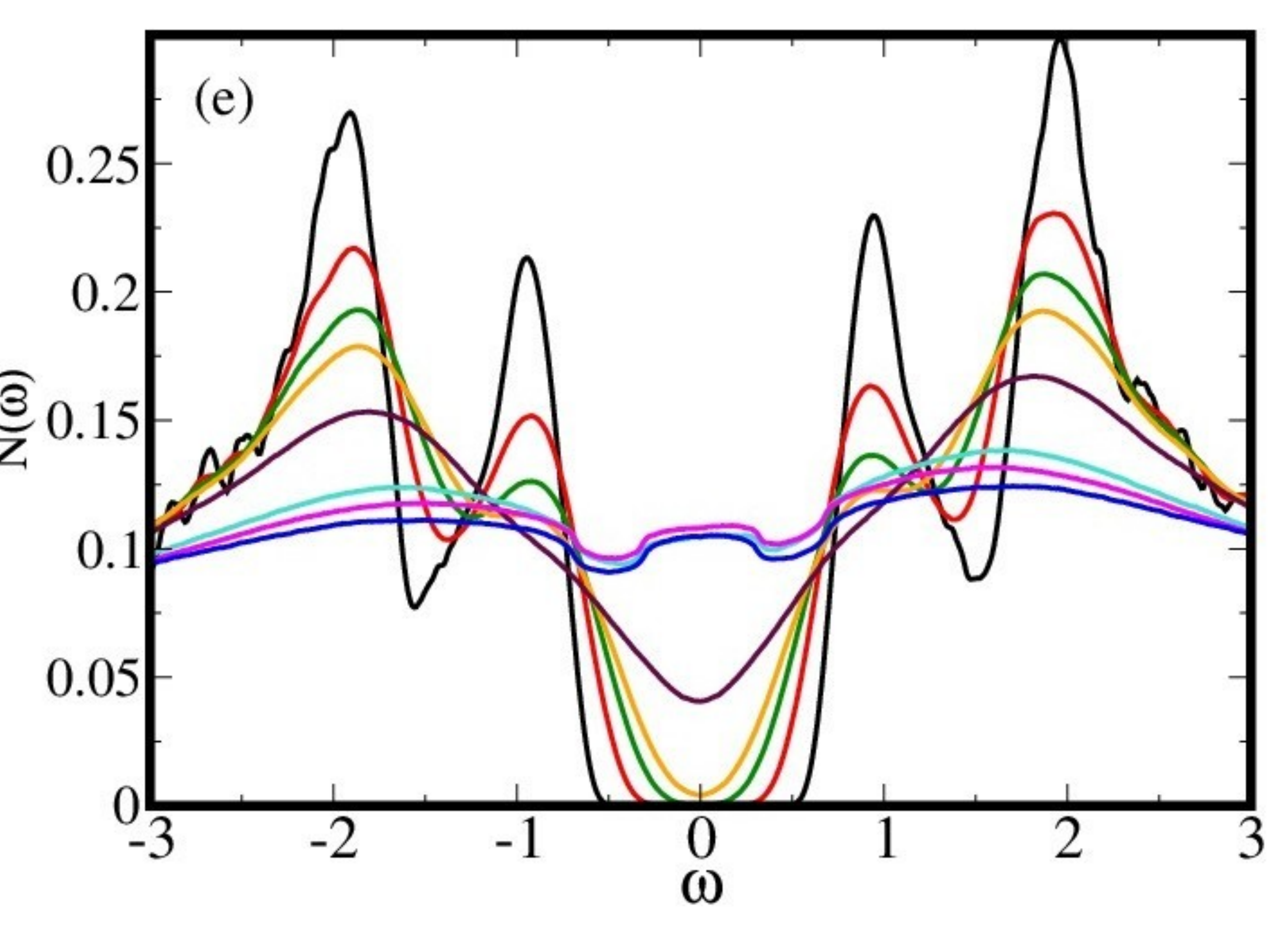}
\includegraphics[width=5.6cm,height=5.0cm,angle=0]{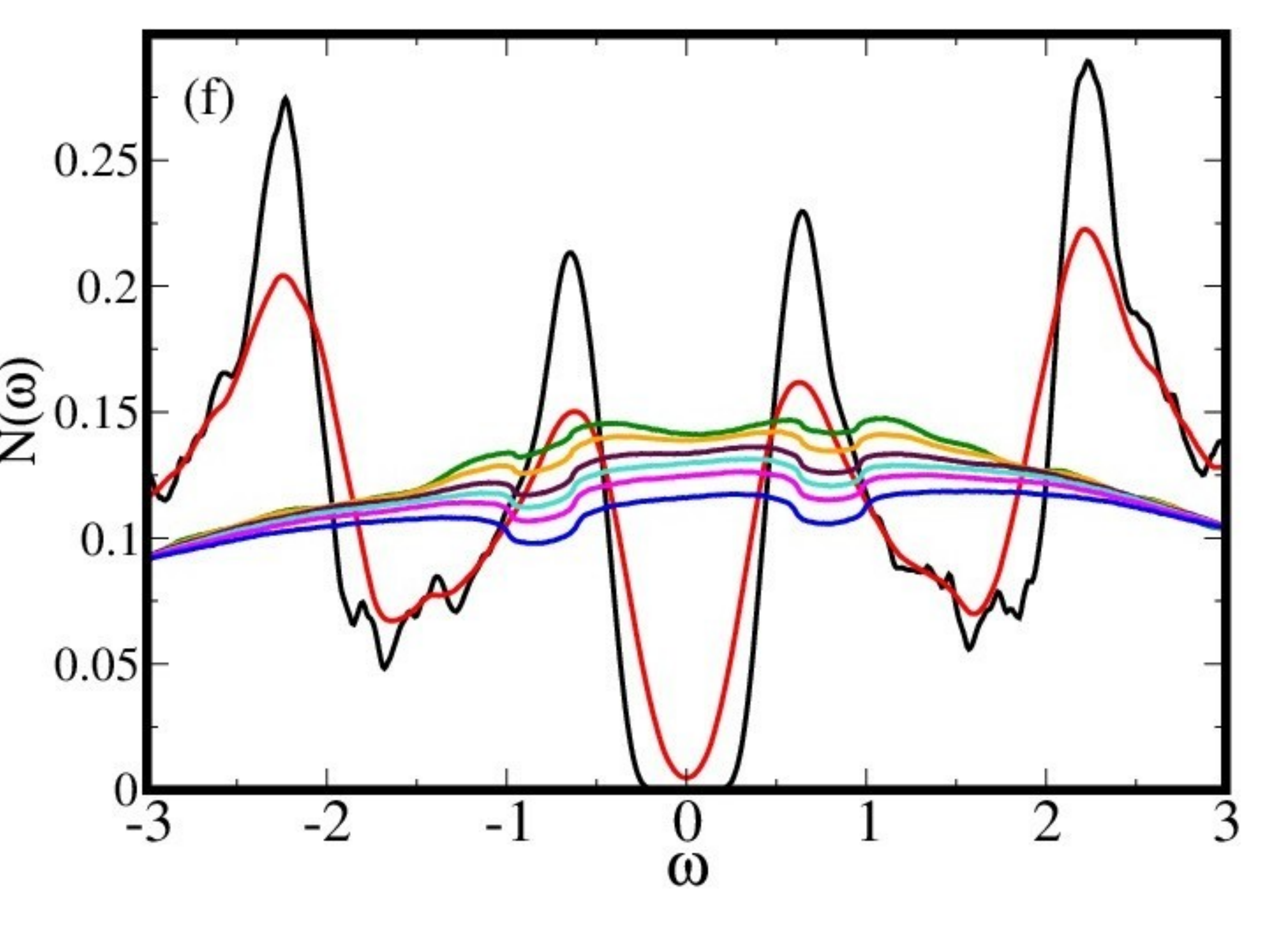}
}
\caption{Color online: Temperature dependence of the DOS 
at $h=0.2t,0.5t,0.8t$
(left to right). The top row shows $N_{\uparrow}(\omega)$ 
and the bottom row
shows the total DOS  $N(\omega)$. $N_{\downarrow}(\omega)$ is
 just a shifted version of
$N_{\uparrow}(\omega)$. While the main feature in (a) is slow
filling up of the gap 
(with increasing $T$ the gap converts to a pseudogap 
already below $T_c$), (b) and (c) reveal that at higher  
fields this `filling up' process is
non monotonic. We quantify the relevant temperature scales later in
Fig.16. }
\end{figure*}

\begin{figure}[b]
\vspace{.5cm}
\centerline{
\includegraphics[width=4.0cm,height=5.0cm,angle=0]{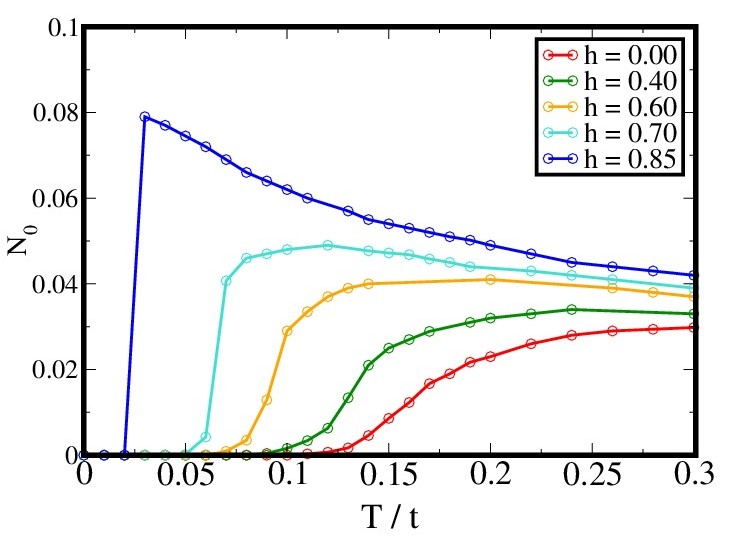}
\hspace{-0.1cm}
\includegraphics[width=4.0cm,height=5.0cm,angle=0]{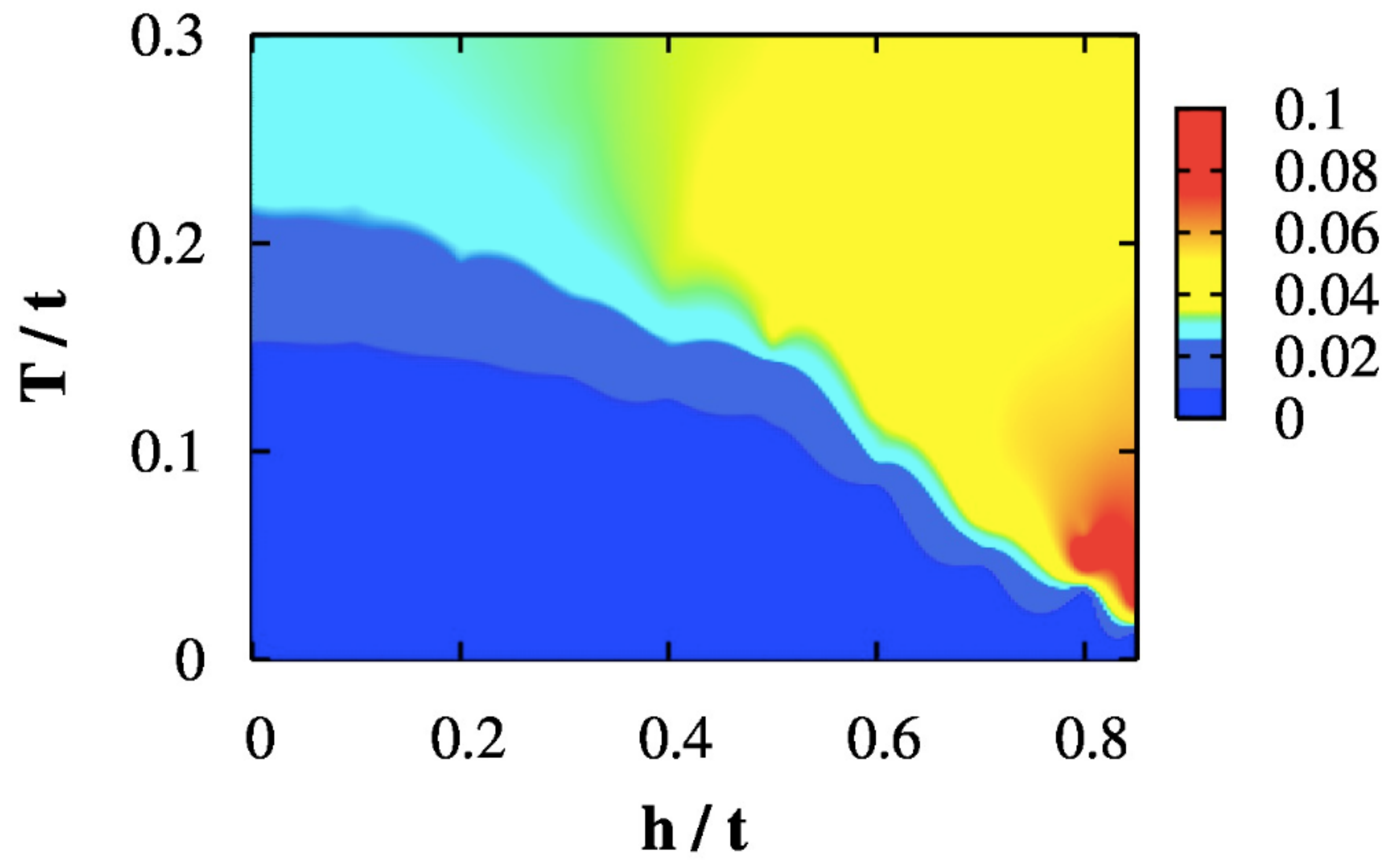}
}
\caption{Color online: (Left) Temperature dependence of the 
DOS at the shifted Fermi level for up spin fermions.
Right: a color map of the DOS at $\omega = -h$ for varying
$h$ and $T$. The non monotonic $T$ dependence is clearly
visible in the large $h$ regime.
}
\end{figure}

\subsection{Density of states}

Pseudogap features in the density of states, and
momentum resolved spectral functions, have
been explored experimentally in the balanced `continuum'
Fermi gas at unitarity. There is also a body of
associated theory
\cite{strinati2012, strinati2002, strinati_nat_phys}.
Unfortunately there are no such detailed
spectral experiments in the
imbalanced case.
Our coupling corresponds roughly to what would be
considered `unitary' (see the Discussion section), but 
we are working on a lattice, at densities far from
the continuum end. To check the usefulness 
of our approach in capturing the qualitative 
features of this well studied end we compared
our `balanced' results to those in the literature
\cite{strinati2012}.
We found that the dispersion and damping share
several features (we will put this up separately)
providing confidence that our lattice results would
have value in analysing {\it imbalanced} continuum
gases as well.

Fig.13 shows the 
spin resolved and total density of states at low,
intermediate, and high fields, $h=0.2t,~0.5t,~0.8t$,
respectively.  We focus on the `up spin' DOS,
since the `down spin' DOS is symmetrically shifted
(and the total is simply a sum of these two) and define
`low energy' as $\omega \sim -h$.

The DOS in the top row in Fig.13 reveal the following
features:
(i)~A hard gap with the usual gap edge 
coherence peaks at low $T$. (ii)~The hard gap converts to a 
pseudogap at a temperature  $T_{pg1}$, where $T_{pg1}(h) 
< T_c(h)$, and the coherence features are 
suppressed with increasing $T$. 
The two features above are visible at all fields,
see panels (a)-(c).
An additional feature is visible in (b) and (c). This is
(iii)~the weakening of the PG
({\it i.e}, increase in the low energy DOS), continues up to a
temperature 
$T_{max}(h)$, beyond which the DOS at low energy {\it falls again}.
The low energy DOS would probably flatten out and the pseudogap 
close at some high temperature.
The somewhat exotic look in the total DOS, panels (d)-(f) 
arises from adding up
$N_{\uparrow}$ and $N_{\downarrow}$ so we focus on
$N_{\uparrow}(\omega)$ itself.

\begin{figure*}[t]
\centerline{
\includegraphics[width=5.5cm,height=5.0cm,angle=0]{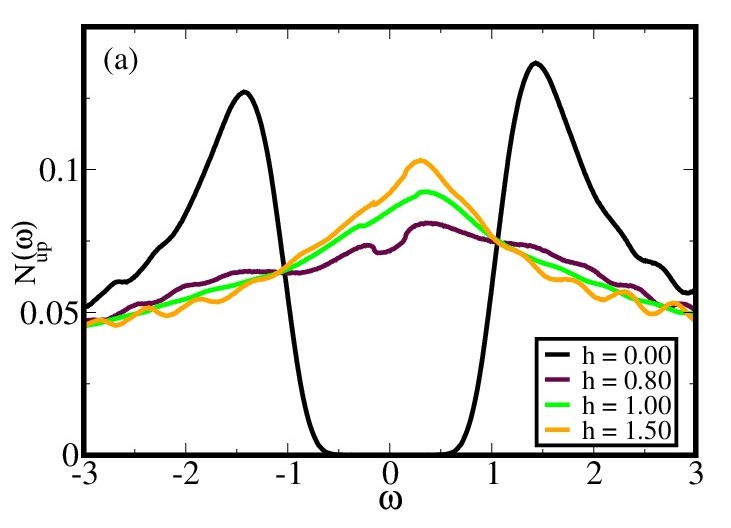}
\includegraphics[width=5.5cm,height=5.0cm,angle=0]{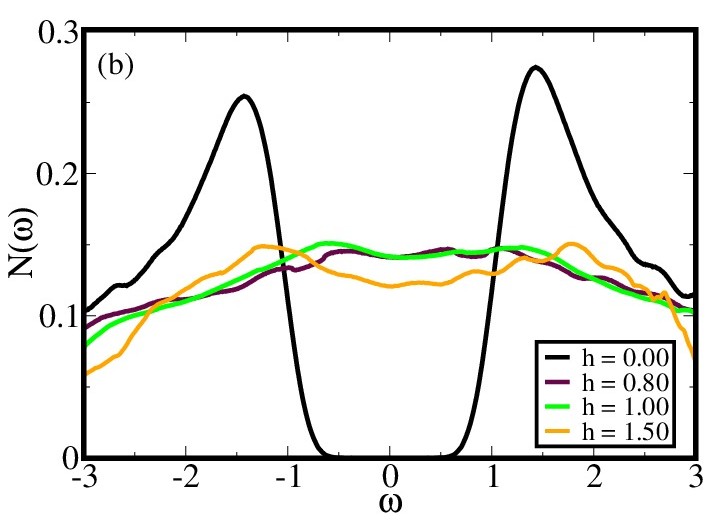}
\includegraphics[width=5.5cm,height=5.0cm,angle=0]{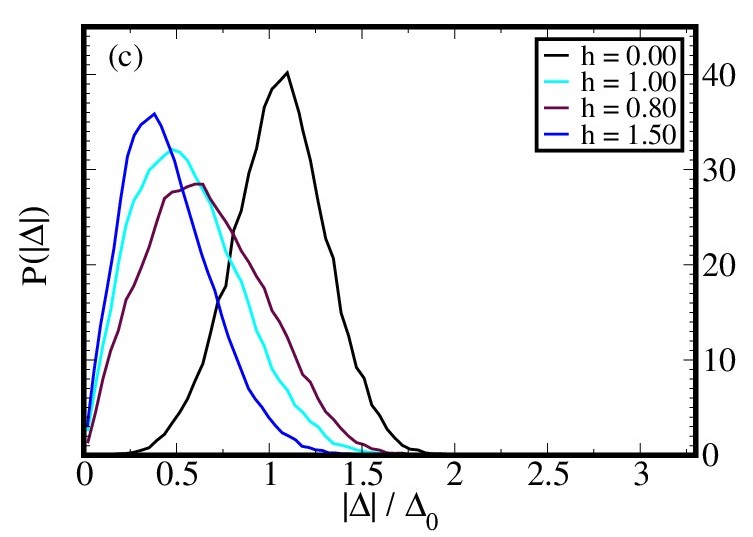}
}
\vspace{.3cm}
\centerline{
\includegraphics[width=5.5cm,height=5.0cm,angle=0]{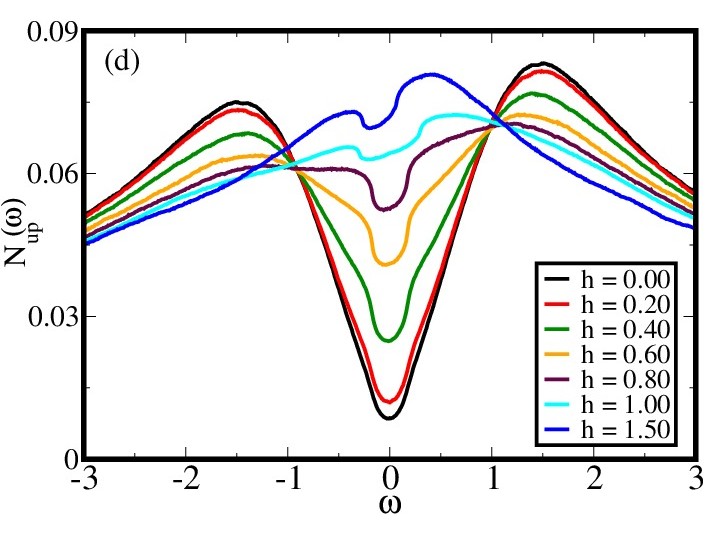}
\includegraphics[width=5.5cm,height=5.0cm,angle=0]{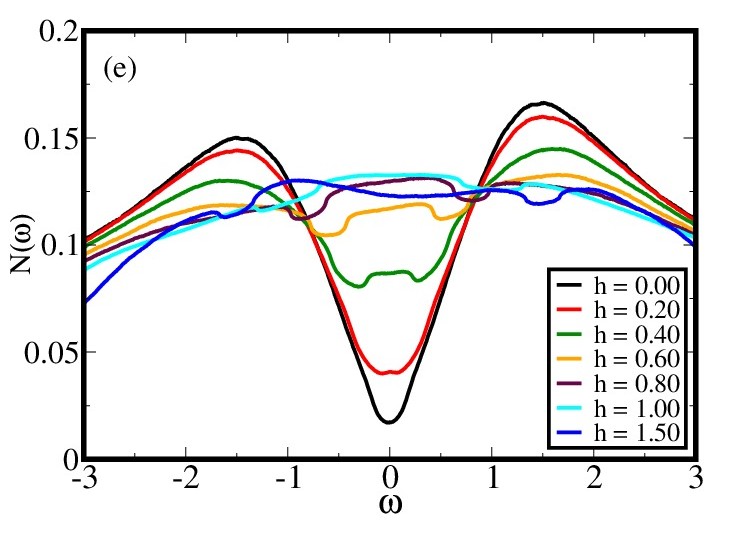}
\includegraphics[width=5.5cm,height=5.0cm,angle=0]{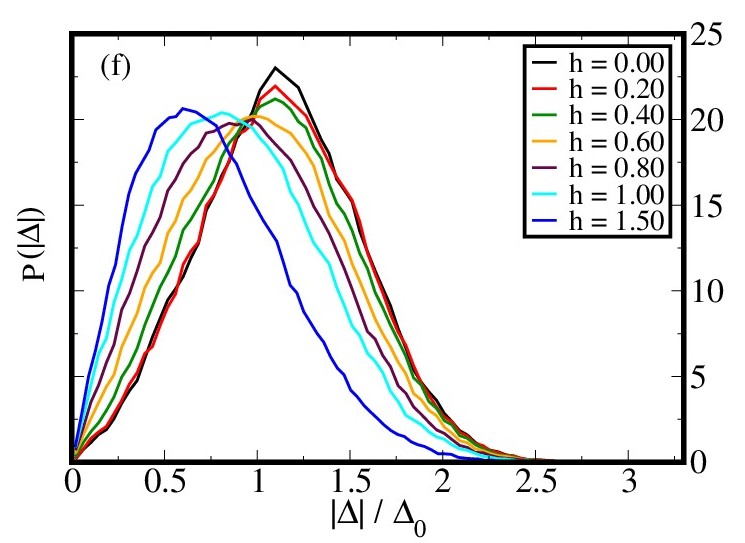}
}
\caption{Color online: Field dependence of the $N_{\uparrow}(\omega+h)$, 
the total DOS $N(\omega)$ and $P(\vert \Delta \vert)$. 
Top row:  $T= 0.05t$, bottom row $T=0.15t$.  
The shift in $N_{\uparrow}$ is to gauge out the field dependent
shift of origin and the resulting clutter in the plot.
}
\end{figure*}

\setcolor \cmykBlack{Fig.14 shows the $T$ dependence of the 
spin up DOS at the up spin  `Fermi level', $\omega=-h$, in the
left panel. }
The right panel shows a map of this density of states as a function
of $h$ and $T$.  These data allow us 
to extract the temperature $T_{max}$ at which the
spin resolved DOS at $\omega= \pm h$ 
has its maximum. 

\begin{figure}[b]
\vspace{.5cm}
\centerline{
\includegraphics[width=6.8cm,height=5.5cm,angle=0]{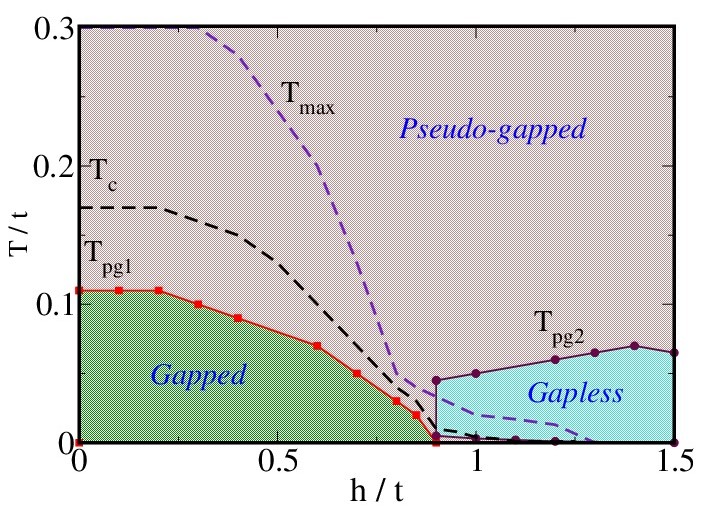}
}
\caption{Color online: Temperature scales associated with the behavior
of the spin resolved 
DOS. The low $T$ hard gap in the superconductor converts to
a pseudogap at a temperature $T_{pg1} < T_c$, while the `ungapped' partially
polarized Fermi liquid at large $h$ {\it develops} a pseudogap
at $T = T_{pg2}$. The entire window above $T_{pg1}$ and $T_{pg2}$ is
pseudogapped. The DOS at the center of the pseudogap ($\omega =  \pm h$)
shows a maximum at $T_{max}$.  }
\end{figure}

Fig.15 shows $h$ dependence at fixed temperatures,
highlighting the effect on the DOS as the system evolves from the
`balanced' situation to the highly magnetized (hence weakly paired)
state.
At $T=0.05t$, top row, which is $\sim 0.3T_c^0$, the pseudogap in the
DOS vanishes at $h \sim t$ while at $T=0.15t$, bottom row, the PG 
seems to persist even at $h = 1.5t$ {\it where the ground state is an
unpaired Fermi liquid!}

In panels (c) and (f) we show $P(\vert \Delta \vert)$ 
for the same field values as in the DOS panels.
At $T = 0.05t$ the $P(\vert \Delta \vert)$ 
remains almost unchanged for  $h$ between $ \sim 0-0.6t$. 
The resulting DOS also remains essentially unchanged over 
this field window.
At $h = 0.8t$ the center of  $P(\vert \Delta \vert)$ 
is at a noticeably smaller value, and as $h$ increases the
peak and the mean value of $\vert \Delta \vert$ shift to
progressively lower value.
At $h=1.5t$ the peak value is $\sim 0.25 \Delta_0$ and the
$P(\vert \Delta \vert)$ cannot generate a pseudogap in
the spectrum.

At  $T = 0.15t$ the 
peak location of $P(\vert \Delta \vert)$ and the mean 
shift to lower value with increasing $h$. 
All the data, except at $h=0,~0.2t$,
are at $T > T_c$.
However, at this temperature the mean value at {\it high
fields} is significantly larger than what we see at $T=0.05t$.
At $h=1.5t$ the peak of $P(\vert \Delta \vert)$ is at 
$\sim 0.5 \Delta_0$, almost twice the $T=0.05t$ value.

As a result, even though the high field system starts at low
$T$ as essentially an `uncorrelated' partially polarized Fermi
liquid (within our scheme) the thermally generated correlation
effects are strong enough to generate a pseudogap with
increasing temperature.  The spin resolved DOS at large
$h$ starts gapless (at $T=0$) but transits to an
interaction induced weak pseudogap phase at high $T$. 
This pseudogap has nothing to do with long range order
in the ground state.

Tracking the field dependence of the pseudogap formation scale,
starting from the low temperature end, allows us to construct
the PG feature based `phase diagram' in Fig.16.
It reveals several intriguing features:
(i)~Although the $T=0$ gap in the spectrum remains the same
for $h=0-0.85t$ the temperature $T_{pg1}$, at which this
gap converts to a pseudogap, collapses as $h \rightarrow
h_{c1}$ the USF-FFLO boundary.
(ii)~Although the mean field ground state has no
pairing for $h \gtrsim 1.3t$, and is therefore gapless, fairly
modest temperature $\sim 0.07t$ generates an weak pseudogap
due to thermal {\it generation of pairing fluctuations}. 
(iii)~The PG in the spin resolved DOS survives to a high
temperature, certainly greater than $T \sim 0.5t$ that we
have probed, although at large $h$ it is a weak feature.
This survival to high $T$ is a consequence of the large
interaction $U=4t$ that we have chosen, and has a parallel
in the PG observations made on the imbalanced cold Fermi gas
at unitarity \cite{ketterle_science_2007}.

\section{Discussion}

Till now we have mainly focused on our specific results. 
In what follows we touch briefly on a few broader issues.
These include:
(a)~the reliability and limitations of our method,
(b)~a conceptual framework for understanding
the numerical data, (c)~the connection between our
intermediate coupling lattice results and the
unitary continuum gas, (d)~qualitative comparison
with cold atom and solid state experiments, and 
(e)~the wider possibilities of our method
in exploring imbalanced superfluids in other situations.

\subsection{Issues of method}

Our results are based on (i)~a Hubbard-Stratonovich decomposition of the
interaction in the pairing channel, (ii)~approximating this auxiliary
pairing field $\Delta_i$ as classical, (iii)~a cluster algorithm 
based Monte Carlo sampling of the $\Delta_i$ field, (iv)~use of 
finite size, as is inevitable in any calculation of this kind.
(ii), (iii) and (iv) introduce errors and we comment on these in the
paragraphs below.  

\subsubsection{Hubbard-Stratonovich decomposition}

The analytic basis of the HS based method is
discussed in the Appendix.

\subsubsection{The static approximation}
The static auxiliary field approximation is exact as $T \rightarrow
\infty$ and in principle becomes less and less accurate as $T \rightarrow 0$
(as the energy difference between the bosonic Matsubara frequencies
reduce). {\it However}, when the ground state has some kind
of long range order, as in both the balanced and unbalanced
fermion cases, the static mode succeeds in capturing much of
the interaction effects. This keeps our $T=0$
results qualitatively valid. A comparison in the balanced
case revealed that by the time $T \sim T_c$, the static
mode captures most of the thermal effects, and anyway for
$T \gg T_c$ it should describe the problem exactly.
Overall, in the current problem, the static approximation
by itself is not a serious limitation.

\subsubsection{Single channel decomposition}

A single field decomposition that is static cannot in
general capture instabilities in all channels.  
In the FFLO regime the pairing, density, and magnetic 
channels are in principle all relevant. However, we 
find that for our chosen mean density, the 
density modulations in the FFLO phase are very weak so 
`density channel' effects are not important (they would
be very important if $n=1$).
The presence of
an additional magnetic channel may make a quantitative 
difference to our results.
While these additional channels are readily incorporated
within MFT a non Gaussian fluctuation theory, like ours,
 involving all these modes is difficult to construct.
We have opted to stay with a simple decomposition 
so that the fluctuation theory can be 
better handled.

{For the homogeneous BP phase we have found that there
is no density
wave ordering tendency at $n=0.94$  and the field regime 
that we have considered. {\it Fluctuations} in density are
present, as evidenced in Figs.7 and 9, but are small
since the field induced magnetization suppresses 
the density wave susceptibility. In a more elaborate
calculation the fluctuations could be numerically
larger, without changing the qualitative features
of our result.}

\subsubsection{Monte Carlo: cluster algorithm and size dependence}

The MC implementation using the Bogoliubov-de Gennes 
scheme requires repeated diagonalisation of the fermion
problem. Done exactly this computation scales as $N^4$ where
$N$ is the system size, limiting one to $N \sim 10 \times 10$,
hardly adequate to access complex phases. This is a primary
limitation in FFLO studies and limits most finite temperature 
studies to mean field theory. We can access much larger
size (up to $40 \times 40$, say) since we use a cluster
based update scheme, discussed in the text. Unfortunately
the cluster size introduces another length scale, that
affects access to FFLO phases, but does not seem to 
have much impact on the uniform SC state. So, as far as
the present study is concerned, size limitations have
not been significant.
We have checked the quality of the MC in the $h=0$
problem earlier by comparing to full QMC \cite{tarat2014}.

\subsection{Landau-Ginzburg framework}

It is useful to put up a Ginzburg-Landau (GL) framework
for qualitatively understanding our results, focusing on
a $\Delta_i$ only theory rather than the `fermion $+~\Delta_i$' 
problem. 
At weak coupling GL theory could have been systematically
derived \cite{casalbouni2004,buzdin1984}, 
here it serves as a phenomenological construct.

The free energy density suggested by
 Casalbuoni {\it et al.} \cite{casalbouni2004}, for 
the superfluid in the presence of a magnetic field, is:
$$
{\cal F} = \frac{1}{2}\alpha \vert \Delta \vert^{2} + 
\frac{1}{4}\beta \vert \Delta \vert^{4} +
\frac{1}{6}\gamma \vert \Delta \vert^{6} + 
\epsilon \vert \nabla \Delta \vert^{2}
+ \frac{\eta}{2} \vert \nabla^{2} \Delta \vert^{2} 
$$
The  complicated form, involving a 6th order
amplitude term and $\nabla^{2} \Delta$, is retained since $\beta$
and $\epsilon$ which are positive in the $h=0$ case can change
sign when $h \neq 0$. 

In the $h=0$ functional involving only $\alpha$, $\beta$ and
$\epsilon$, we have $\beta > 0$. The sign change of $\alpha$
drives a {\it second order transition}
 to a ${\bf q} = (0,0)$ state since the
gradient term penalizes spatial modulation.

$\beta$ changing sign from positive to negative leads to a first
order transition, again to an uniform state if $\epsilon > 0$,
and one retains a positive $\gamma$. On the other hand if
$\epsilon$ changes sign the system would head towards a modulated
state, whose wave number has to be decided by the presence of a
positive $\eta$. This would be the thermal transition to some
FFLO state.

In the continuum weak coupling limit it turns out that $\beta$ and
$\epsilon$ change sign from positive to negative {\it at the same 
point}\cite{casalbouni2004,buzdin1984}. 
In that situation one has a second order normal to SC transition at
weak field, crossing over to a first order normal to FFLO transition
beyond a critical field.
 
Our lattice 
mean field results at $U=4t$ indicate that a first order thermal
transition need not be necessarily to an FFLO state. We do have
a window of a first order normal to uniform SC transition. This
distinction is probably a lattice versus continuum difference.
It shows up in the MC results as well, with $T_c$ scales suppressed
due to amplitude and phase fluctuations.

The MC results suggest the rough behavior of the various GL
coefficients in terms of $h$ and $T$ but the observed 
metastability is harder to pin down. For that a 
more elaborate functional, derived by tracing out
the fermions from the coupled problem, and expanded about
${\bf q} =0$ and ${\bf q} = {\bf Q}$ (the FFLO wave vector)
would be needed.
For $h < h_1$
that free energy has the deepest minimum at ${\bf q} =0$,
and also uniquely reaches this state on thermal 
cycling. For $h_1 < h < h_{c1}$ the absolute minimum is
still at ${\bf q} =0$ but it seems 
that the minimum at  ${\bf Q}$, although metastable,
dominates the energy landscape.
When one cools from the high $T$ state
the system seems to first encounter this ${\bf q} \neq 0$ 
minimum and tracks this state down to $T=0$.

\subsection{Connection to continuum unitary gas}

While we have motivated our lattice model
in terms of experiments on the continuum unitary 
Fermi gas there are 
issues which need highlighting.
These are (i)~the notion of `unitarity' in the
2D lattice model (and its relation to BCS-BEC
crossover), and (ii)~the access to continuum `universal'
effects via lattice simulations.

{\bf Unitarity:}
The primary scale quantifying the strength of interaction
in cold Fermi gases is the two body $s$-wave scattering length: 
$a_{D}$, where $D$ denotes the spatial dimensionality.
The dimensionless coupling constant can then be written in terms
of $k_Fa_{D}$, where $k_F$ is the Fermi wavevector. 
We quickly comment on unitarity and the BCS-BEC crossover 
issue in the continuum and lattice contexts and then see what
information lattice simulations can yield.
 
In the  3D continuum Fermi gas the scattering length 
$a_{3D} \rightarrow \infty$ as the interaction
$g \rightarrow g_c$. $g_c$ is a finite in 3D, and defines
the interaction for which a  
two body bound state first forms in vacuum.
The (inverse) dimensionless coupling $1/k_Fa_{3D} =0$ at $g_c$.
This is also the point near which the transition temperature
of the 3D Fermi gas has its maximum, with $T_c^{max}/E_F \sim 0.15$.
On the  3D Hubbard lattice an  equivalent critical interaction
for bound state formation can be worked out and yields
$U_c/t \sim 7.9$. Again, the $T_c$ is found to be maximum 
for $U/t \sim 8$ (from QMC), remarkably close 
\cite{beck} to $U_c$. 

In the 2D continuum a two body bound state forms in 
the presence of an arbitrary
attractive interaction so $a_{2D}(g) \rightarrow \infty$ for
$g \rightarrow 0$. This is however in the deep BCS regime
where a weak coupling description in terms of fermionic
quasiparticles is sufficient. As $g$ increases the pair
size shrinks and in the Bose limit $a_{2D} \rightarrow 0$.
The  crossover coupling in the 2D case is defined 
via $ln(k_Fa_{2D}) \rightarrow 0$, {\it i.e}, the scattering
length being comparable to inter particle separation. 
We are not aware of a 2D continuum QMC calculation for the
$T_c$, but interpolation between the BCS and BEC end suggests
that the maximum $T_c$ occurs for $ln(k_Fa_{2D}) \rightarrow 0$,
with \cite{ries2015} $T_c^{max}/E_F \sim 0.1$.

On the 
2D Hubbard lattice a two body bound state would form at
arbitrary weak attraction, {\it i.e}, $U_c/t
\rightarrow 0$. The notion of $k_F$ is not very meaningful
on the lattice,
particularly away from low density, but following the
cases above one may identify the crossover as the region of
maximum $T_c$. The $T_c$ is well established via QMC 
and the maximum 
occurs for $U/t \sim 5$. For $U/t=4$
that we use the $T_c$ is $\sim 0.9$ of the maximum value \cite{paiva}.

Overall, the two body bound state based
unitary point in 3D {\it also} corresponds to a regime where 
(i)~neither
the fermionic nor the bosonic quasiparticle description
suffices, and (ii)~the $T_c/E_F$ is maximum. In 2D the
simple two body argument would put this regime at extreme
weak coupling but (i) and (ii) above indicate that 
$[ln(k_Fa_{2D})]^{-1} \rightarrow \infty$, rather than
$a_{2D} \rightarrow \infty$, is the relevant choice.

Our interaction strength is not far from what would be considered the
`unitary' value in the 2D lattice case. Can we comment on the 
universal physics one would have seen in the continuum case,
on which, remarkably, there is now an experiment?

{\bf Continuum universality from lattice physics:}
The first difficulty is with our density choice, we have used 
$n \sim 0.9$ to maximize $T_c$ (at the same time avoiding the
density wave instability at $n=1$). At this density the lattice
effects are very prominent and the Fermi surface is distinctly
non circular. Even if we had used lower density, $n \sim 0.1$,
say, where the Fermi surface is indeed circular and
$\epsilon_{\bf k} \sim k^2$ is a good approximation access
to continuum effects is difficult. If the
interaction were {\it weak}, {\it i.e}, $U \ll 4t$, the 
physics would have been insensitive to the high energy band cutoff.
However, `unitarity' requires $U \sim 5t$ and even if the Fermi level
is at the lower edge of the band, scattering effects couple in states
at the upper edge. The high energy states are lattice specific, and
as the paper by Privitera {\it et al} \cite{castellini2010, castellini2012}
demonstrates even at reasonably low densities one obtain results that do not match
with continuum predictions. So, extracting `universal' physics
from lattice simulations require extremely low densities, $\sim 0.001$,
or lattice sizes $\sim 10^4$, outside the range of what is doable today. 

\begin{figure}[t]
\vspace{.5cm}
\centerline{
\includegraphics[width=4.0cm,height=5.0cm,angle=0]{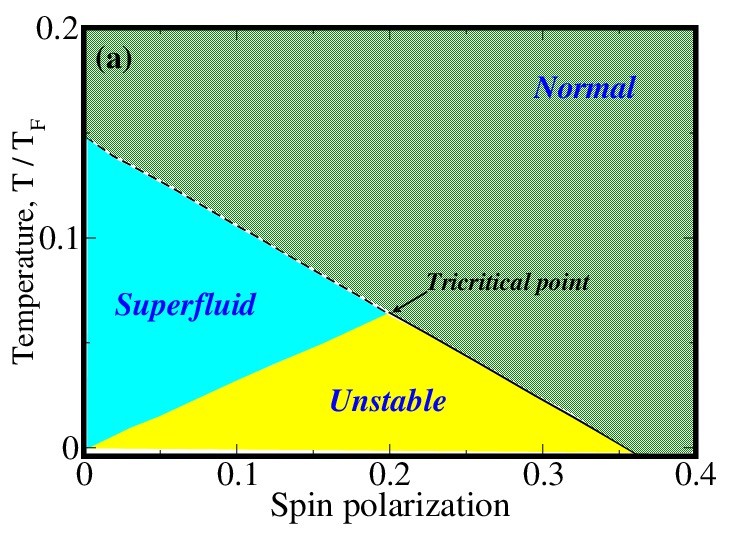}
\includegraphics[width=4.0cm,height=5.0cm,angle=0]{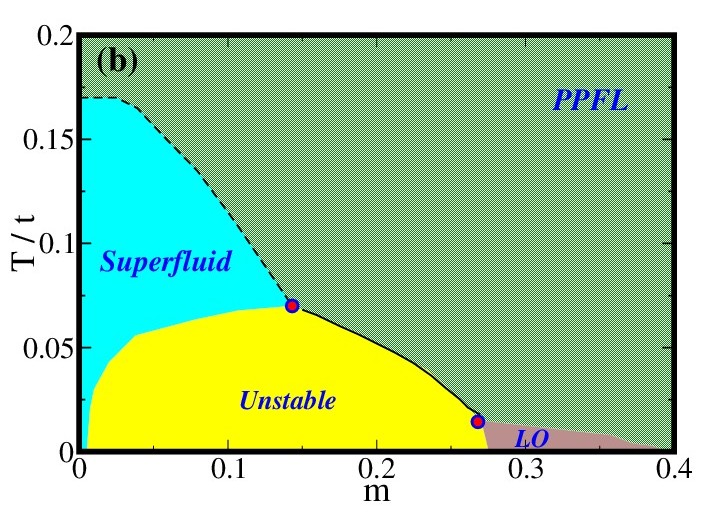}
}
\caption{Color online: The imbalance-temperature
phase diagram inferred from measurements on a cold atomic
gas at unitarity (left) \cite{ketterle2008},
compared to our result on the intermediate
coupling (peak $T_c$) Hubbard model (right).
The normalization of the $x$ axis is same in both panels, while
the $y$ axis have different reference scales.
}
\end{figure}

\subsection{Comparison with experiments}

\subsubsection{Atomic superfluids}
 
Our model finds it most appropriate experimental
counterpart in imbalanced cold Fermi gases, studied at
strong attractive interaction promoting s-wave pairing.
There are still differences, 
{\it e.g}, (i)~our results are on a lattice theory, 
the experiments are in the `continuum',
(ii)~there is a trap present in the experiments, 
and (iii)~dimensionality
(the experiments are in three dimensions).
Nevertheless, the similarities are striking.
Fig.17 presents the experimental phase diagram \cite{ketterle2008} of the
unitary Fermi gas in terms of magnetization
$m = (n_{\uparrow} - n_{\downarrow})/(n_{\uparrow} + n_{\downarrow})$
and temperature, constructed by the experimental group by 
`gauging out' the effect of the trap.
Next to it we show our $m-T$ phase diagram, constructed at
$n \sim 0.9$ by varying $h$ (hence $m$). 

The experiments infer (i)~a homogeneous magnetized superfluid (SF),
(ii)~an unstable region, and (iii)~a  magnetized normal Fermi liquid.
At $T=0$ the SF to unstable transition occurs at $m=0+$, suggesting
that a finite $m$ homogeneous SF state cannot occur at $T=0$, while
the `unstable' to normal transition occurs at $m \sim 0.35$.
The $T_c$ of the unitary Fermi gas is significantly lowered with
respect to mean field theory \cite{torma2007} 
and using the measured zero imbalance $T_c$ as the reference 
scale, the tricritical point occurs at $m \sim 0.2$ and $T_{tri}
\sim 0.4 T_{c}^0$.
{The suppression of $T_{c}$ in presence of imbalance is an
extension of the zero field (balanced) behavior \cite{haussmann2007, 
bulgac2007, burovski2007, akkineni2007}. The presence of imbalance 
makes the suppression rapid.
}

The $m-T$ picture that emerges from our data already has the
fluctuation effects built in on $T_c^0$. In the ground state
the SF to `unstable' transition occurs at  $m=0+$ and the unstable
to LO transition at $m \sim 0.28$ and the LO to normal transition 
at $m \sim 0.37$.
{It must be noted that the unstable region essentially is a 
phase separated region, marked by discontinuity in density,
and characterized by the absence of homogeneous superfluid phase.
}
The LO state has not been observed experimentally
in the 2D geometry, possibly due to additional fluctuations in
the absence of a lattice.
Our tricritical point is at $m \sim 0.15$ and $T_{tri} \sim 0.07t
\sim 0.4 T_c^0$.
Given that we are in a regime where the Fermi surface is significantly
non circular the overall correspondence of phase boundaries and
temperature scales is reasonable. We have predictions about
spectral properties on the $m-T$ plane that we will present
separately.

\subsubsection{Superconductors}

The solid state systems in which Pauli limited behavior
is observed, for example CeCoIn$_5$ and the organic
superconductors, are non s-wave materials and have `low energy'
fermionic degrees of freedom even in the ordered
state. In our model, however, the fermions are gapped,
or have a strong pseudogap, due to the large on site
attraction - unless the population imbalance is large.
As a result, despite the overall similarity in the look
of our theory phase diagram and those observed in experiments
\cite{bianchi2003,tayama2002,lortz2007},
 the comparison of indicators like specific heat, $C_V(T,h)$,
and magnetization, $m(T,h)$, reveal differences in
detailed behavior. We have made these comparisons but
do not present the data here.

\subsection{Extensions of the present method}

The present work was focused on understanding a 
part of a larger phase diagram. As a natural extension of this we have
studied the thermal properties of the large $h$ FFLO states in
detail. We have also computed the momentum resolved spectral functions 
of the BP, PPFL, and FFLO phases over the entire $h-T$ window. We will 
present these results separately.

A natural extension of the present method, involving a `two field'
decomposition, can handle the effect of disorder  
\cite{tarat2_2014,randeria,cui_fflo_dis} on the FFLO
state, including the thermal effects which are in general difficult
to access. 

Finally, cold Fermi gases involve a trapping potential and a non
trivial spatial dependence of the region where the fluid is
magnetized. While experimental optical lattice sizes $\sim 100 \times
100$ are hard to access using our MC technique, we hope to access the
physics at least in the BP regime using a local density scheme grafted
on to our Monte Carlo solver.

\section{Conclusion}

We have used a real space Monte Carlo technique based
on a static pairing field approximation 
to study the behavior of a Pauli limited superconductor in the 
BCS to BEC crossover regime.
We find that the $T_c$ scales are strongly suppressed with
respect to mean field predictions, there is a wide window
of metastable FFLO states in which the system gets trapped
when the true ground state is a homogeneous superfluid, 
and the spin resolved density of states shows a non monotonic
low energy character.
We do not know of Pauli limited solid state systems
with s-wave pairing, but ultracold unitary 
gases suggest
an universal phase diagram quite similar to what we observe. 
This paper probes the lower field `breached pair' state in
detail, companion papers discuss the FFLO regime and 
the spectral features expected with changing imbalance.

{\it Acknowledgments:}
We acknowledge discussions with J. K. Bhattacharjee and 
use of the High Performance Computing Cluster at HRI. 
PM acknowledges support from an 
Outstanding Research Investigator Award of the DAE-SRC.

\section{Appendix: Hubbard-Stratonovich 
transformation and Monte Carlo sampling}

The primary numerical technique we use is a Monte Carlo
implementation of a `single channel' static auxiliary
field decomposition of the A2DHM \cite{tarat2014, meir_nature,
tarat2_2014, evenson1970}.
Below we discuss about the various aspects of our numerical
technique.

The Hubbard model at strong interaction requires a non perturbative
solution. The exponential growth in the dimension of the Hilbert
space rules out the use of exact diagonalization except for very
small sizes. The `exact' tool of choice is quantum Monte Carlo (QMC)
against which all approximations are bench marked. While QMC can be
implemented via various approaches, the method below easiest reveals
the connection to our approach.

The Hubbard partition function is written as a functional integral
over Grassmann fields $\psi_{i \sigma} (\tau),
{\bar \psi}_{i \sigma}(\tau)$.
\begin{eqnarray}
Z & = &  \int {\cal D} \psi {\cal D} {\bar \psi}
e^{-S[ \psi,  {\bar \psi}]} \cr
S & = & \int_0^{\beta} d\tau
[
\sum_{ij,\sigma} 
\{ {\bar \psi}_{i\sigma} 
(\partial_{\tau} \delta_{ij}  + t_{ij}) 
\psi_{j\sigma} \}
- \vert U \vert \sum_i 
{\bar \psi}_{i\uparrow} {\psi}_{i\uparrow}
{\bar \psi}_{i\downarrow} {\psi}_{i\downarrow}
]
\nonumber
\end{eqnarray} 
Only quadratic path integrals can be exactly evaluated. Since 
the interaction generates a quartic term in the $\psi$'s the 
partition function cannot be immediately evaluated.

The quartic term is `decoupled' exactly through a Hubbard-Stratonovich
transformation in terms of pairing fields $\Delta_i(\tau),{\bar
\Delta}_i(\tau)$.  This introduces a term
$\Delta_i {\bar \psi}_{i\uparrow}(\tau) {\bar \psi}_{i\downarrow}(\tau)$
in the action. 
\begin{eqnarray}
Z & = & \int  {\cal D} \Delta {\cal D} \Delta^* {\cal D} \psi {\cal D} {\bar \psi} 
e^{ - S_1[ \psi,  {\bar \psi}, \Delta, \Delta^*]} \cr
S_1 & = & \int_0^{\beta} d\tau
[
\sum_{ij,\sigma}
\{ {\bar \psi}_{i\sigma}
(\partial_{\tau} \delta_{ij}  + t_{ij})
\psi_{j\sigma} \} 
\nonumber \\ && + \sum_i \{\Delta_i(\tau) {\bar \psi}_{i \uparrow}(\tau)
{\bar \psi}_{i \downarrow}(\tau)
+ h.c + { \vert \Delta_i\vert^2 \over \vert U \vert } \} ]
\nonumber
\end{eqnarray}
The $\psi$ integral is now quadratic but an
additional integration over the field $\Delta_i(\tau)$ has been
introduced.
The `weight factor' for the  $\Delta$ configurations can be determined by
integrating out the  $\psi,{\bar \psi}$, and using these weighted 
configurations one goes back and computes fermionic properties.
Formally
\begin{eqnarray}
Z &=&  \int  {\cal D} \Delta {\cal D} \Delta^* 
e^{-S_2 [ \Delta, \Delta^*]} \cr
S_2 & = & log[Det[{\cal G}^{-1} - \Delta]] + {\vert \Delta_i\vert^2 \over \vert U \vert}
\nonumber
\end{eqnarray}
where ${\cal G}$ is the Greens function associated with the non interacting $H$.

The weight factor for an arbitrary space-time configuration $\Delta_i(\tau)$
involves computation of the fermionic determinant in that background.
If we write the auxiliary field $\Delta_i(\tau)$ in terms of its Matsubara
modes, as $\Delta_i(\Omega_n)$, then the various approximations can be
readily recognized and compared.

\begin{itemize}
\item{Quantum Monte Carlo retains the full `$i,\Omega_n$' dependence of $\Delta$
computing $log[Det[{\cal G}^{-1} - \Delta]]$ iteratively for importance
sampling. The approach is valid at all $T$, but does not readily yield
real frequency spectra.}
\item{Mean field theory restricts $\Delta_i(\Omega_n)$ 
to a spatially uniform (or periodic)
and time independent ($\Omega_n=0$) mode, {\it i.e},
$\Delta_i(i\Omega_n) \rightarrow \Delta$. The 
free energy is minimized with respect $\Delta$. 
When the MF order parameter vanishes at high temperature 
the theory trivializes.}
\item{Our static auxiliary field (SAF) approach retains the full spatial dependence
in $\Delta$ but keeps only the $\Omega_n=0$ mode,
{\it i.e}, $\Delta_i(\Omega_n) \rightarrow \Delta_i$. 
It thus includes 
classical fluctuations of arbitrary magnitude but no quantum ($\Omega_n \neq 0$)
fluctuations. One may
consider different temperature regimes: (1)~$T=0$: since classical 
fluctuations die off at $T=0$, SAF reduces to standard Bogoliubov-de Gennes (BdG) 
MFT. (2)~At $T \neq 0$ we consider not
just the saddle point configuration but {\it all configurations} following
the weight $e^{-S_2}$ above. These involve the classical amplitude and
phase fluctuations of the order parameter, and the BdG equations are solved
{\it in all these configurations} to compute the thermally averaged
properties. This approach suppresses the order much
quicker than in MFT. (3)~High $T$: since the $\Omega_n=0$ mode 
dominates the exact partition function the SAF approach  
becomes exact as $T \rightarrow \infty$.}
\item{DMFT: for completeness we mention that DMFT retains the full dynamics
but keeps $\Delta$ at  effectively one site, {\it i.e},
$\Delta_i(\Omega_n) \rightarrow \Delta(\Omega_n)$.}
\end{itemize}

Overall, our method reduces to BdG mean field theory only at $T=0$ but
retains all the classical thermal fluctuations at $T \neq 0$. 
As a result it is only as good as MFT at $T=0$ but is far superior
in estimating $T_c$, and essentially exact as $T \rightarrow \infty$.
It does use BdG iteratively as a tool but on all fluctuating configurations
not just the mean field state.



\begin{thebibliography}{99}
\bibitem{var-comment}
In contrast to Chiesa and Zhang, who do a mean field theory in 
pairing, density, and spin channel, we decouple the Hubbard term
only in the pairing channel.  This would lead to different results
in parts of the FFLO regime but has no effect on the USF ground
state.
\end{thebibliography}
\end{document}